\documentclass{fac-arxiv}

\usepackage{lineno,hyperref}
\modulolinenumbers[5]

\usepackage{amsfonts,amstext,amsmath,amssymb}
\usepackage[dvipdfmx]{graphicx} 
\usepackage{bmpsize}
\usepackage{oz}           
\usepackage{xcolor}
\usepackage{fancyhdr}

\usepackage{stfloats}
\fnbelowfloat 

\newcommand{\wmmRef}{weak-memory trace refinement\ }
\newcommand{\WMMRef}{Weak-memory trace refinement\ }

\newcommand{\PEv}{{\cal{P}}rogEvents}
\newcommand{\ObjEv}{{\cal{O}}bjEvents}

\newcommand{\obs}{{eff}}

\newcommand{\sem}[1]{\lbag #1 \rbag}
\newcommand{\semM}[1]{\lbag #1 \rbag_M}

\newcommand{\order}[1]{\ <_{#1}\,}
\newcommand{\Porder}{\ <_{P_M}\,}

\newcommand{\res}[2]{#1\,_{|#2}\,}  
\newcommand{\resIR}[1]{#1\,_{|ir}\,}  
\newcommand{\resP}[1]{#1\,_{|p}\,}  
\newcommand{\obsT}[1]{#1\,_{|global}\,}

\newcommand{\linM}{~{\sf lin}_M~}

\newcommand{\refsto}{\sqsubseteq}

\renewcommand{\proof}{\noindent{\bf Proof}}

\newcounter{axiom}
\newcounter{soundprops}
\newcounter{transprops}

\newcounter{compprops}
\newcounter{comprops}

\renewenvironment{proof}{\par\noindent\textbf{Proof}\\}{\vspace*{-2ex}\\\hspace*{\textwidth}\hspace*{-6ex}$\Box$}
\newenvironment{myproof}[1]{\par\noindent\textbf{Proof}~#1\\}{\vspace*{-2ex}\\\hspace*{\textwidth}\hspace*{-6ex}$\Box$}
\newtheorem{theorem}{Theorem}
\newtheorem{lemma}{Lemma}
\newtheorem{definition}{Definition}

\title{A sound and complete definition of linearizability\\ on weak memory models}
 \author{G. Smith, K. Winter and R. Colvin\\\smallskip
 School of Information Technology and Electrical Engineering,\\
             The University of Queensland, Australia
}

\begin{document}

\maketitle

\begin{abstract}
Linearizability is a widely accepted notion of correctness for concurrent objects. Recent research has investigated redefining linearizability for particular hardware weak memory models, in particular for TSO. In this paper, we provide an overview of this research and show that such redefinitions of linearizability are not required: under an interpretation of specification behaviour which abstracts from weak memory effects, the standard definition of linearizability is sound and complete on {\em all\/} hardware weak memory models. We prove our result with respect to a definition of object refinement which takes a weak memory model as a parameter. 
The main consequence of our findings is that we can leverage the range of existing techniques and tools for standard linearizability when verifying concurrent objects running on hardware weak memory models. 

\end{abstract}

\section{Introduction}
\label{sec:introduction}

Linearizability \cite{HeWi90} is a widely accepted notion of correctness for concurrent objects \cite{Moi07} that relates
the behaviours of an object's implementation to the possible behaviours of its specification.
As a correctness notion it benefits greatly from being \emph{compositional}, i.e., 
the linearizability of each object of a system in isolation guarantees that the
overall system is also linearizable. This provides us with a practical approach to 
proving correctness.

At the level of implementation, operations on an object take time and hence they may overlap in a multi-threaded program. This is obviously difficult to reason about. Linearizability allows us to prove, however, that the behaviour of such an object implementation is consistent with that of a specification in which operations are atomic, and hence cannot overlap. The key concept is the notion of a {\em linearization point\/}, a point where an operation in the implementation can be thought of as taking effect atomically. Choosing such a point for each operation in a concurrent history of an implementation allows us to match that history with a sequential history of the specification. The sequential history is referred to as a {\em linearization\/} of the concurrent history.

 \begin{figure}
 \centering
 \scalebox{.45}{\includegraphics{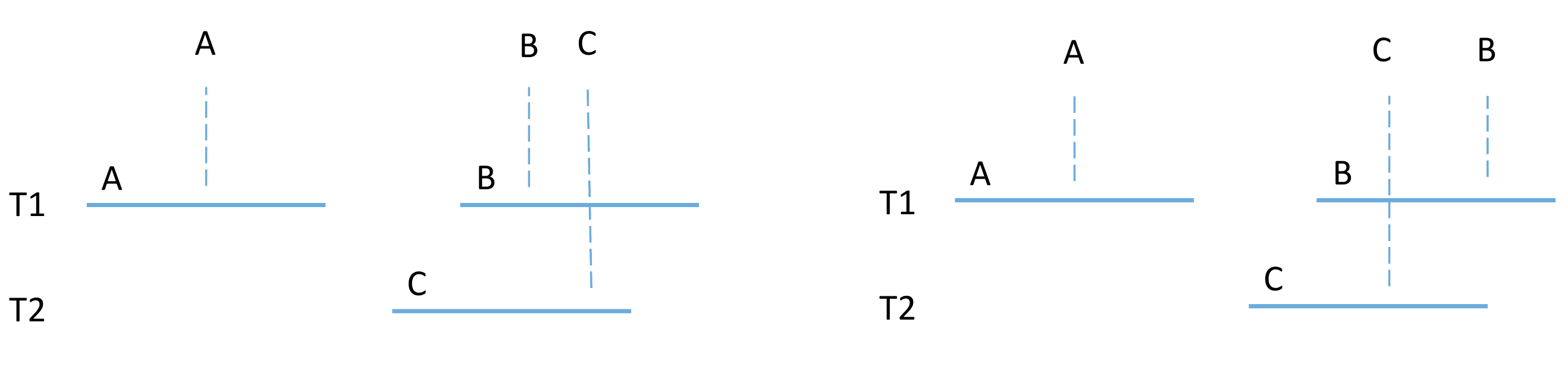}}
 \caption{Linearizability example}
 \label{fig:lin}
 \end{figure}

Figure~\ref{fig:lin} shows two linearizations of a concurrent history in which operation B of thread T1 overlaps with operation C of thread T2. There are two important things to note. Firstly, the linearization point of an operation must occur somewhere between its {\em invocation}, i.e., when it is called, and its {\em response}, i.e., when it returns. This means that operations which do not overlap in the concurrent history occur in the linearization in the order they were invoked. Secondly, overlapping operations in the concurrent history may occur in either order in the linearization, regardless of the order of their invocations and responses. A concurrent history satisfies a specification, provided one of its possible linearizations is a history of the specification. 

A concurrent object is considered correct when each of its {\em finite\/} histories linearizes with a sequential history of the specification. Hence, linearizability only checks safety properties of an object implementation, not liveness properties \cite{GotsmanYang2011,Smith2017}. Since linearizability is compositional, we can prove the correctness of a system of interacting objects by showing each component object is linearizable with respect to its specification \cite{HeWi90}. 

Recent work \cite{bur12,got12,DBLP:conf/hvc/TravkinMW13,ifm,DohertyDerrick2016, DongolVMCAI2018, DohertyIFM2018} examines the applicability of linearizability in the context of weak memory models of modern multicore architectures \cite{Sewell:2010:XRU:1785414.1785443,UnderstandingPOWER,AxiomaticPOWER,Alglave:2009:SPA:1481839.1481842,HerdingCats,armv8,ColvinFM2018}. These memory models improve hardware efficiency by reducing accesses to global memory. Individual threads may operate on local copies of global variables, updates to the global memory being made by the hardware and largely out of the programmer's control.\footnote{A programmer can add fences (or memory barriers) to code to force any pending updates to be written to memory. However, if used indiscriminately, fences reduce the gains in efficiency that a weak memory model provides.} This can cause threads executing on different cores to get out of sync with respect to the values of global variables. 

For example, on the TSO (Total Store Order) architecture \cite{Sewell:2010:XRU:1785414.1785443}  a thread updating a global variable $x$ stores the new value in a per-core FIFO buffer. Threads executing on that core will then read $x$ from the buffer, rather than the global memory, until the new value is flushed from the buffer by the hardware. In the meantime, threads on other cores read the value of $x$ from the global memory or from their own core's buffer when it has a value for $x$. \medskip

There have been several attempts at defining linearizability for TSO. Burckhardt et al.\ \cite{bur12} include a notion of buffers in the specification of a concurrent object, and associate two atomic steps with each specification operation: one where the effect of the operation updates the buffer, and a subsequent one where it takes effect in the global memory. Gotsman et al.\ \cite{got12} introduce nondeterminism into the specification to model that a thread may, or may not, have seen a recent update. Both of these approaches change the specification that the implementation needs to satisfy. The resulting specifications are less intuitive and do not correspond to specifications that would normally be found as part of a software library. 

\begin{figure}
 \centering
 \scalebox{.45}{\includegraphics{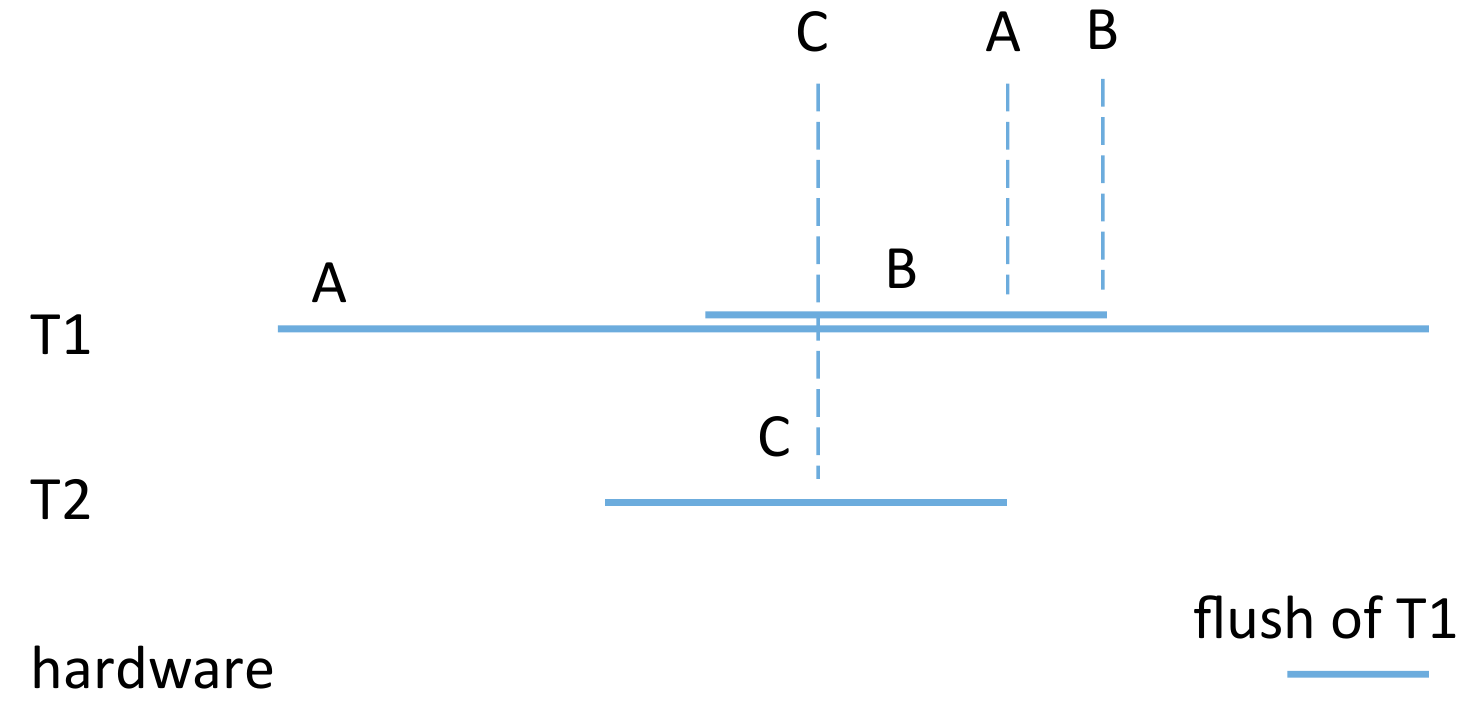}}
 \caption{Linearizability example on TSO}
 \label{fig:lin-tso}
 \end{figure}

Derrick et al.\ \cite{ifm} take a different approach, leaving the specification unchanged and instead changing the definition of linearizability. A similar approach is used by Travkin et al.\ \cite{DBLP:conf/hvc/TravkinMW13,hvc14}  when developing tool support for proving linearizability on TSO. These approaches (which we will collectively call {\em TSO-linearizability\/}) change the bounds within which the linearization point must occur. Specifically, they do not require that the linearization point of an operation occurs before its response; it can occur any time up to the final flush associated with the operation. This could allow operation A of Figure~\ref{fig:lin} to linearize after C in the case when its flush occurs after C (see Figure~\ref{fig:lin-tso}). 

TSO-linearizability is the basis for the proof methods in \cite{der14} and \cite{Derrick2017}. As proved in \cite{DBLP:conf/fm/DerrickS15} it is compositional. 
It takes a particular view on the meaning of a specification in which, although operations occur atomically, they do not necessarily take effect immediately. 
Assume, for example, we have an object $o$ with operations $WriteX$ which writes value 1 to a variable $x$, $WriteY$ which writes value 1 to a variable $y$, and $ReadX$ and $ReadY$ which read the variables $x$ and $y$ respectively. When a straightforward implementation of $o$'s operations (i.e., one not involving fences) is run on TSO, the program in Figure~\ref{fig:unsound} could result in both $z$ and $w$ being equal to 0 (the initial value of both variables). This occurs when the flushes of both writes are delayed until after both reads have executed. This somewhat surprising behaviour is considered correct by TSO-linearizability since there is a 
linearization as shown in Figure~\ref{fig:tso-bad}.

\begin{figure}
  \vspace{-8mm}
 \[
   \qquad \qquad\qquad \qquad\qquad\qquad\qquad~~~~\parbox[t]{2.1cm}{~~{\sf T1} \vspace*{2mm}\\ {\sf o.WriteX;\\ z=o.ReadY}} 
     \parbox[h]{1cm}{\vspace*{1.5cm}$\zBig\parallel$} 
      \parbox[t]{3cm}{~~{\sf T2}\vspace*{2mm}\\ {\sf o.WriteY;\\ w=o.ReadX}}
 \]
 \caption{Program that can result in $z=w=0$ on TSO}
  \label{fig:unsound}
 \end{figure}

While this is a valid view of correctness, it requires the user to have a solid understanding of the memory model in order to know what behaviour the object might engage in. In particular, as can be seen in Figure~\ref{fig:tso-bad}, operations may effectively occur out-of-order with respect to {\em program order\/}, i.e., the order they appear in the program text. Under simple memory models like TSO, this is not too much of a burden for the user, but may become so under more complex memory models, especially when the code of the object also becomes more complex.

\begin{figure}
 \centering
 \scalebox{.45}{\includegraphics{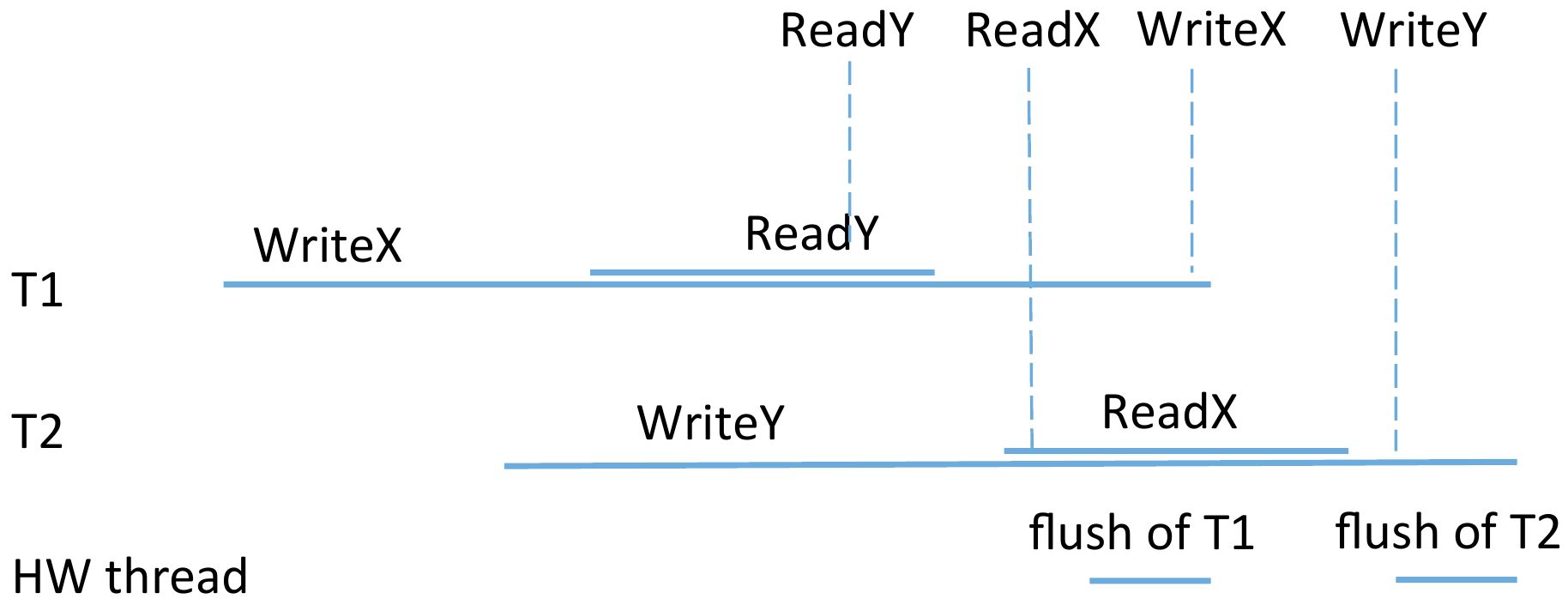}}
 \caption{Linearization of the program of Figure~\ref{fig:unsound}}
 \label{fig:tso-bad}
 \end{figure}

Strengthening the definition of TSO-linearizability to ensure that events take effect in program order is proposed in \cite{DBLP:conf/fm/DerrickS15} and \cite{Dongol2015a}. Under these definitions, the program in Figure~\ref{fig:unsound} should never result in both $z$ and $w$ equal to 0 if the operations behave according to their specification.
While this 
overcomes the above problem, it introduces a different one: the definition is no longer compositional. To see this, consider the example in Figure~\ref{fig:comp-fails} where $o1$ and $o2$ are different objects with an operation $Write$ (to write a local variable of the object) and an operation $Read$ (to read the local variable).

\begin{figure}
\vspace{-8mm}
 \[
   \qquad\qquad\qquad\qquad~~\parbox[t]{2.5cm}{~~{\sf T1} \vspace*{2mm}\\ {\sf o1.Write(1);}} 
     \parbox[h]{1cm}{\vspace*{1.5cm}$\zBig\parallel$} 
      \parbox[t]{3cm}{~~{\sf T2}\vspace*{2mm}\\ {\sf o2.Write(1);\\ x=o1.Read( )}}
      \parbox[h]{1cm}{\vspace*{1.5cm}$\zBig\parallel$} 
      \parbox[t]{5cm}{~~{\sf T3}\vspace*{2mm}\\ {\sf while(z = 0) z=o1.Read( );\\ y=o2.Read( )}}
 \]
 \caption{Compositionality counter-example (triangular race)}
  \label{fig:comp-fails}
 \end{figure}

Assume the implementation of each object (when run in isolation) is linearizable. When the objects are run together, compositionality would ensure that the combined system is also linearizable. However, the composed system of Figure~\ref{fig:comp-fails} can result in both $x$ and $y$ being set to 0 on TSO. This is possible since $T2$ may perform its write to $o2$ and read from $o1$ before $T1$ writes to $o1$ (thus setting $x$ to 0), but not flush the value it wrote to $o2$ until after the other two threads have run to completion (setting $y$ to 0). 

This outcome is not possible, however, according to a specification based on operations taking effect in program order: if $x$ is set to 0 then this step and hence the write to $o2$ on $T2$ must have taken effect before the write to $o1$ on $T1$ took effect, and hence before the read of $o2$ on $T3$. Hence, the composed system is not linearizable. 

Doherty and Derrick \cite{DohertyDerrick2016} provide a variant of TSO-linearizability which preserves program order and is proved compositional when the client program is restricted to be free from \emph{operation races} (like data races but at the level of operations rather than individual lines of code). 
The program in Figure~\ref{fig:comp-fails} provides an example of such a racy program showing a {\em triangular race\/} \cite{DBLP:conf/ecoop/Owens10} and hence Doherty and Derrick's approach is not applicable. While their result is a useful contribution, it does not allow us to prove linearizability of objects which may be used in {\em any\/} context: something we would like for objects in a software library.\medskip

What is central to the quest to defining a suitable notion of linearizabilty is a reference point
for correctness under weak memory models that allows one to prove
soundness and completeness. What does it mean for an object implementation to behave
correctly, and which behaviour of an object can be deemed incorrect?
The answer to this seemingly simple question requires some attention
in the context of weak memory models. In \cite{refine18} we
provided such a notion of correctness, \emph{object refinement},
which is based on a notion of trace refinement where the object is viewed in the context of a client program.
This notion is similar to contextual or observational refinement \cite{Dongol2016,Filipovic-LinvsRef2010}
but is geared for the context of weak memory models.
Object refinement is
parameterised by the memory model it refers to, and is therefore
generic and can be instantiated for any hardware weak memory model. 
 In this work we utilise this notion to provide a formal
proof of soundness and completeness.  In contrast to all previous
work, we prove that using standard linearizability (as originally
proposed by Herlihy and Wing \cite{HeWi90}) is sound and complete
not just on TSO, but {\em any\/} hardware weak memory model. It is, of
course, also compositional (as proved by Herlihy and Wing).

To do this, we adopt an alternative view of a specification in which its operations occur {\em and take effect\/} atomically (rather than potentially taking effect later). This view of specifications has two advantages. Firstly, it ensures that an implementation maintains the \emph{intent} of the specification, e.g., an implementation of a lock operation does not have a delayed effect that allows multiple threads to acquire the associated lock \cite{ifm}.
Secondly, understanding a linearizable object's specified behaviour within a program does not require a knowledge of the memory model.
Analogously to a specification of a concurrent object abstracting from the effects of interleaving, 
the specification in this view also abstracts from the  effects of weak memory models.

The paper is structured as follows.
In Section~\ref{sec:prog} we outline the basic concepts of our theory
including that of \emph{effects}\footnote{Referred to as {\em observations\/} in our earlier work \cite{refine18}.} which is key to our definitions. 
Based on these concepts, Section~\ref{sec:progsemantics} formalises the semantics of 
programs under a given memory model and postulates some basic axioms that we assume of program 
behaviour under weak memory.  
The semantics of a concurrent object in the context of a client program, under
a memory model, is elaborated in Section~\ref{sec:objectsemantics},
distinguishing cases for the specification, in which operations are
atomic, and the implementation which includes non-atomic, and possibly
non-terminating, operations.
Section~\ref{sec:wmmref} ties these basics into the notion of \emph{\wmmRef}$\!\!$ which defines
refinement under memory model $M$ for a client program using an object and its specification, respectively.
The definition is parameterised with a given memory model for which we assume
a semantics is given. 
Using \wmmRef we then define our notion of \emph{object refinement} under memory model $M$
which delivers the notion of correctness for objects.
In Section~\ref{sec:lin}, we prove that the standard definition of linearizability is both sound and complete with respect to our definition of object refinement. 
Section~\ref{sec:ex} illustrates how
 object refinement, and hence standard linearizability, can be used to prove correctness of typical concurrent objects.
The paper concludes with a discussion of related work in Section~\ref{sec:relwork}.\medskip

\noindent{\bf Contributions.} This paper is an extension of our previous paper \cite{refine18} in which our notion of object refinement was first defined. That paper provided a reference point from which definitions of linearizability on different hardware weak memory models could be proved sound and complete. The definition of object refinement abstracts from the semantics of particular memory models, instead relying on an operational semantics, such as those for TSO \cite{Sewell:2010:XRU:1785414.1785443} and the significantly weaker memory models POWER and ARM \cite{armv8,ColvinFM2018}, to provide possible program behaviours. As it explicitly avoids out-of-thin-air results, it is not applicable to software weak memory models such as C11 \cite{BattyOSSW11,NienhuisMS16}.

In this paper, we use that parameterised definition to show that standard linearizability is sound and complete for {\em all} hardware weak memory models. 
The assumptions that are required for the proofs are made explicit as axioms of the program semantics. These axioms are derivable from concepts, like object encapsulation and  atomicity, and are independent of weak memory model behaviour.  
This is in contrast to previous work which has changed the definition of linearizability, often for a specific memory model. Earlier in this section, we have provided an overview of this body of earlier work and related the different definitions.

The main consequence of our findings is that we can leverage the range of existing techniques and tools for standard linearizability when verifying concurrent objects running on hardware weak memory models. 

\section{Client programs and effect events}
\label{sec:prog}

To investigate the behaviour of concurrent objects under hardware weak memory models, 
and relate their implementations to their specifications using refinement, 
we need to consider the calling context.
Programs calling the operations of a concurrent object are referred to as \emph{client programs},
or clients for short. A client program $P$ is concurrent, running multiple threads $T_i$ possibly 
on multiple cores, and is affected by the memory model of the architecture it is running on. 
For some finite $n$, we have\footnote{For simplicity, we do not consider dynamically spawned threads.}
\[ P \sdef T_1 \parallel T_2 \parallel \ldots \parallel T_n .\]

Following other work on concurrent objects \cite{HeWi90, Filipovic-LinvsRef2010, Dongol2016}, the behaviour of a program is
described in terms of \emph{events} that occur. We allow events to be \emph{program steps}, \emph{operation events} or, as introduced in Section~\ref{sec:observations}, \emph{effect events}.

\emph{Program steps} are any steps performed by the client other than calling an object operation.
These are assignments, conditional branch instructions (e.g., {\sf if} 
or {\sf while} statements), other control instructions like various forms of fences,
atomic read-modify-write instructions which atomically perform these three steps 
(e.g., the compare-and-swap instruction CAS),
and higher-level instructions which can, in many cases, be defined in
terms of assignments and/or conditional branches. For example, a statement
{\sf await(z=1)} could be defined as {\sf while(z $\neq$ 1)\,$\{\}$}.

\emph{Operation events} abstract the behaviour of an operation call by a program. They include the \emph{invocation} of the operation (i.e., when it is called) and
the operation's \emph{response} (i.e., when it returns). The operation events carry the operation's input and output values
as parameters and thus reflect the operation's externally visible behaviour. The internal 
behaviour of an object is elided.  



\subsection{Effect events}
\label{sec:observations}

Central to our definitions is
the notion of an {\em effect event\/}. 
Such an event denotes the point in an execution where 
a program step or an operation 
can be deemed to have taken effect and (if observable) has
been observed by \emph{all} threads $T_i$ of client $P$. 
Effect events are where the results of program steps and output values of operations (if any) become known globally.

Note that an operation's internal steps and their effects are not observable on the program level (due to object encapsulation), and hence
we record only one effect event per operation call. 
These operation effects are not used for object refinement. Object refinement is based on the refinement of the client program and
only takes into account observable program steps.
However, operation effects are essential when defining the semantics of
objects in a context of a client running under a weak memory model, 
as we will see in Section~\ref{sec:objectsemantics}.

Introducing effect events allows us to decouple when events occur in a program
and when they are observed by other threads, which might not fall together under hardware weak memory models. 
Program steps and invocation and response events 
correspond to the points in an execution where the program counter ({\sf pc}) is at the address of the corresponding instruction. For an invocation or response, this instruction will be an update to {\sf pc}, setting it to where the execution should continue. Hence, for any memory model these events will remain in the order that they occur in the program text. 

The semantics specific to the memory model under consideration
determines the possible orderings of the effect events. 
Effect events occur only after their matching program event or, in the case of an operation, after the matching invocation.
Furthermore, for any program step or operation that writes to memory, its effect will occur when \emph{all} threads can either

\begin{enumerate}
\item[(a)] access the new value of a global program variable written by a program step, or
\item[(b)] access the values of \emph{all} shared object variables written by an operation.
\end{enumerate}


%
%

\subsection{Programs}
\label{sec:prog-ex}

The behaviour of a concurrent program $P$ under a weak memory model $M$ can be described 
by a partial order $\Porder$ over the events of the program to which all possible executions of that
program must adhere. When event $e$ occurs before event $e'$ in the partial order, i.e., $(e,e') \mem \Porder$ then event $e$ must occur before event $e'$ in any execution of $P$. When $P$ involves branching (e.g., {\sf if} or {\sf while} statements), there will be several paths through $P$ (one corresponding to each combination of branches). The events of different paths need to be distinct which can be ensured, for example, by including in each event an identifier of the path to which it belongs.  

 \begin{figure}
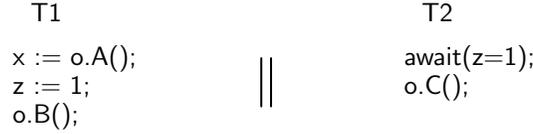

   \vspace{-8ex}
 \[
   \qquad \qquad\qquad\qquad\qquad\qquad\parbox[t]{3.2cm}{~~{\sf T1} \vspace*{2mm}\\ 
      {\sf x := o.A();\\ z := 1;\\ o.B();}} 
     \parbox[h]{2cm}{\vspace*{1.8cm}{\resizebox{!}{0.5cm}{$\parallel$}}} 
      \parbox[t]{3cm}{~~{\sf T2}\vspace*{2mm}\\ {\sf await(z=1);\\ o.C();}}
 \]
 \caption{Client program}
  \label{fig:client}
 \end{figure}

To understand what a partial order of a program for a particular memory model might look like
consider the example of a client program given in Figure~\ref{fig:client}
with two global program variables {\sf x} and {\sf z}, running on two threads {\sf T1} and {\sf T2}. 
The threads call three operations of the same object {\sf o}, of which the operation {\sf B} does not write
to any shared object variable.
 Note that we assume that the result of any operation call is implicitly 
 stored in a local register (e.g., {\sf r}$_{\sf A}$ for the operation call
 {\sf o.A}), and ${\sf inv_A}$ denotes the invocation of {\sf o.A}, etc.

 On a {\em sequentially consistent\/} (SC) architecture, i.e., one
 without a weak memory model, writes to global variables occur
 instantaneously. Hence, the effect event for a program step
 occurs immediately after the program step, and that of an operation
between its invocation and response.
 The partial order on the events of the program of
 Figure~\ref{fig:client} on SC is shown below. The branch in the partial order after the program step {\sf z=1} of thread {\sf T1} takes effect corresponds to the operation call {\sf o.B()} of {\sf T1} occurring in parallel with the code of thread {\sf T2}.
 
\begin{center}
 \scalebox{.5}{\includegraphics[viewport=4 663 565 832]{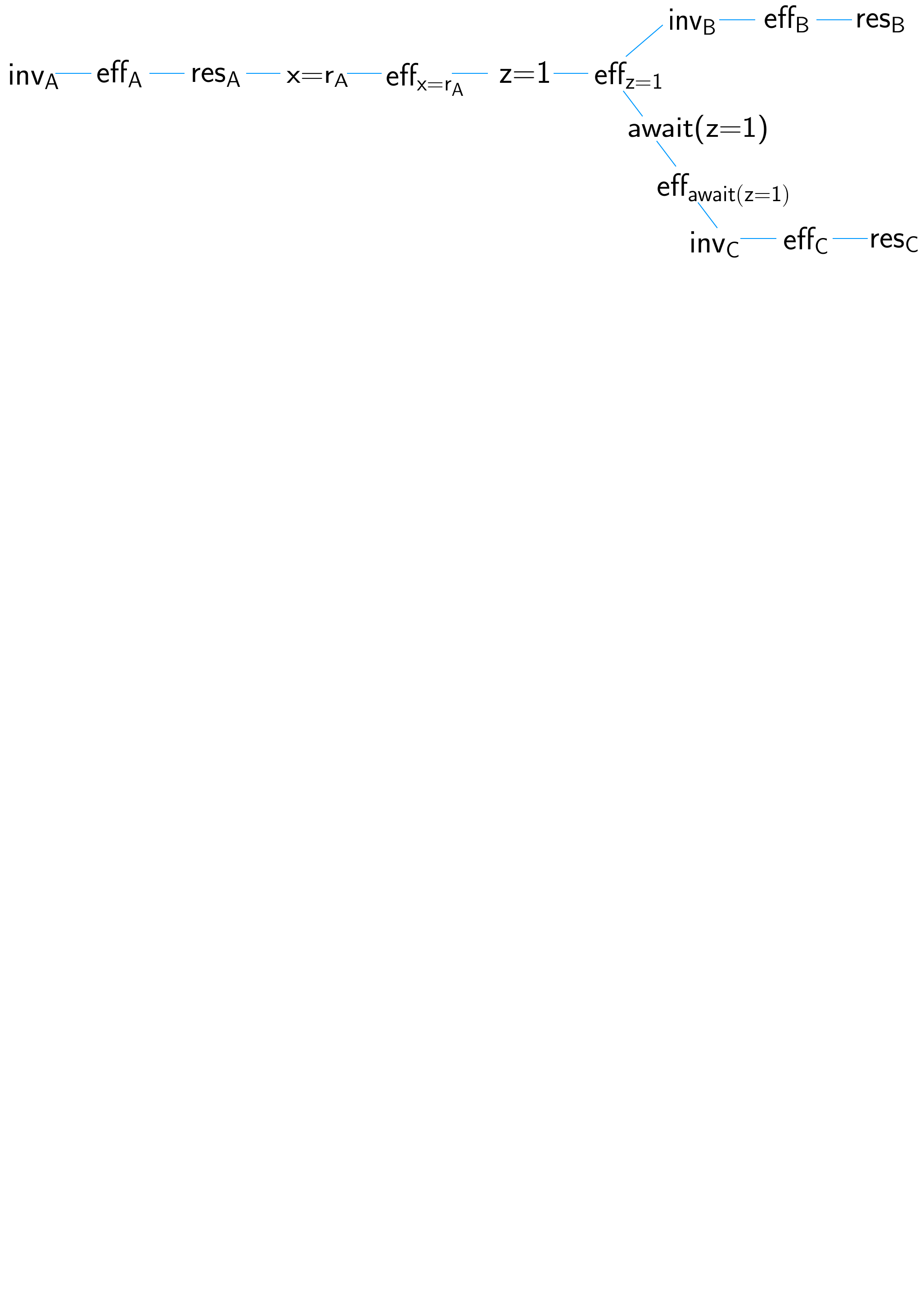}}
\end{center}

On TSO \cite{Sewell:2010:XRU:1785414.1785443}, writes to global variables and shared object variables 
become available to threads on other cores when they are flushed. 
These flushes may occur at some later point, but we know that they occur in the same order as the writes occurred.
Hence, in the partial order of the events on TSO (depicted below),
${\sf \obs_{A}}$ occurs before ${\sf \obs_{x=rA}}$ which occurs before
${\sf \obs_{z=1}}$. The effect event ${\sf \obs_{B}}$ can only occur
after ${\sf inv_B}$ but is not ordered with respect to the other effect events since
{\sf o.B} does not write to any shared variable, and hence does not have to follow 
the FIFO order of flushes of writes.
Similarly, the effect events ${\sf \obs_{await(z=1)}}$ and ${\sf \obs_{C}}$ are  un-ordered. 

\begin{center}
 \scalebox{.5}{\includegraphics[viewport=4 664 505 797]{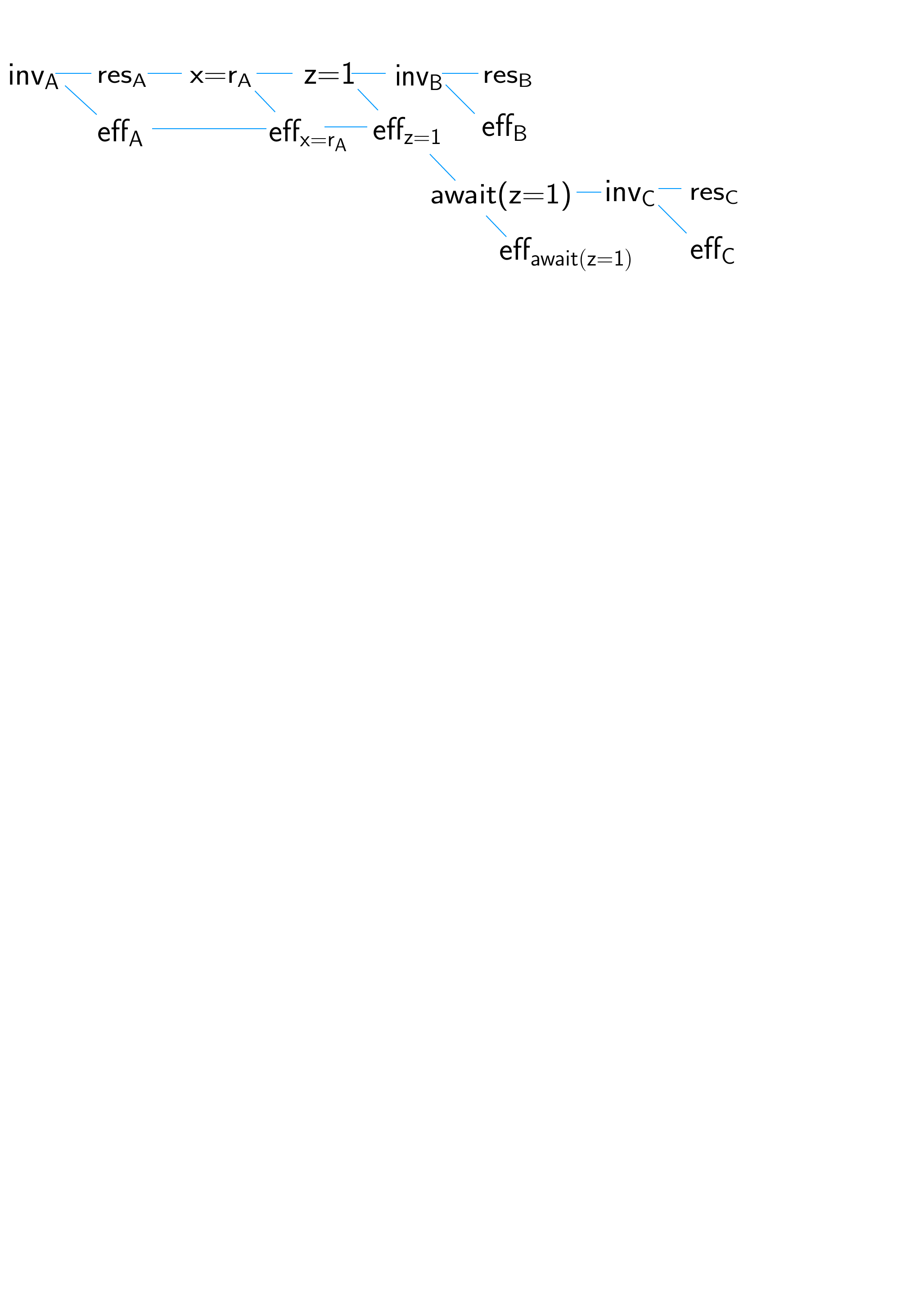}}
\end{center}

On ARM and POWER \cite{UnderstandingPOWER,AxiomaticPOWER,Alglave:2009:SPA:1481839.1481842,HerdingCats,armv8,ColvinFM2018}, the effect events might occur out of (program-) order
and also flushing of variables need not occur in a FIFO manner.
Hence the effects are un-ordered in the partial order of the
program depicted below, apart from ${\sf \obs_A}$ and ${\sf \obs_{x=rA}}$ which are ordered
due to a data-dependency on register ${\sf r_A}$.
As under TSO, the assignment to {\sf z}  and  the {\sf await} statement maintain their order
due to the data dependency (on {\sf z}) and
the synchronisation ({\sf await}) between {\sf T1} and {\sf T2}. 


\begin{center}
 \scalebox{.5}{\includegraphics[viewport=4 656 477 798]{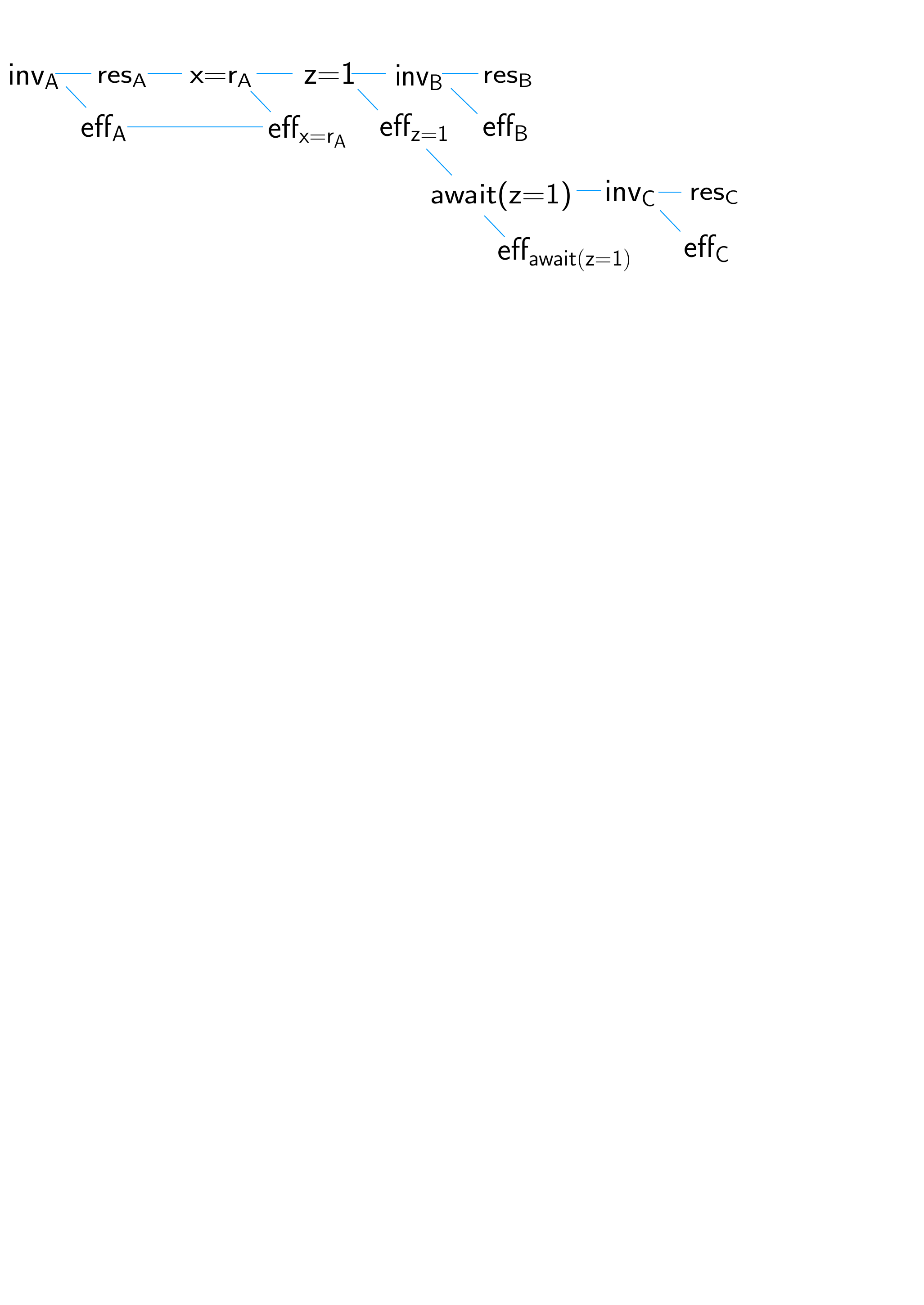}}
\end{center}

For programs with more than two threads, the semantics of ARM and POWER also allow {\em non-multi-copy atiomicity\/}, i.e., where threads on different cores to the thread performing a write may see the write at different times. On such a memory model, the effect will be the point where {\em all\/} cores have seen the write. In the operational semantics of ARM and POWER in \cite{ColvinFM2018}, non-multi-copy atomicity is modelled using a sequence of writes where each write is associated with the set of thread identifiers of threads which have seen the write. The effect occurs in this semantics when that sequence indicates that all threads have seen either the write or a {\em future\/} write to the same variable, i.e., one that occurs later in the execution.\medskip


The order in which effect events occur in an actual behaviour of a program determines the value 
of the output associated with the effect. This value must also correspond to the output value of the operation response
or program step that is observed through the effect event. This correspondence, which prevents out-of-thin air results available in software memory models such as C11 \cite{BattyOSSW11,NienhuisMS16}, is captured in a well-formedness
condition on traces formalised in the next section.

\section{Semantics of programs relative to the memory model}
\label{sec:progsemantics}

The semantics of a program $P$ under  memory model $M$ is defined in terms of
the set of {\em events\/} that can occur, and a partial order $\!\!\Porder$ over those events. 
%
%
The events are, on one hand, determined by $P$ through the
program text, and, on the other hand, determined by 
  the semantics of the memory model as outlined in Section~\ref{sec:prog}. 
In this section, we formalise the semantics of a program assuming its events and partial order are given. 

\subsection{Events}

Let $T$ be the set of all thread identifiers, and $Call$ the set of all (unique)
operation calls. An operation is then defined as a call by a
particular thread.

\[Op\sdef T\cross Call\]

Let $PS$ denote the set of all program step events, and $Val$ the set
of all values (of input and output parameters), including a special
element $\bot$ meaning `no value'. The set of all events is defined as
follows, where each invocation is associated with an input, and each
response and effect with an output.

\[Event  \sdef  step(T,PS) | \obs(T,PS) | inv(Op,Val) | res(Op,Val) | \obs(Op, Val) \]
In the remainder of this paper, we refer to $step(T,PS)$ and $\obs(T,PS)$ as program events, $\PEv$, 
and to $inv(Op,Val)$, $res(Op,Val)$, and $\obs(Op, Val)$ as object events, $\ObjEv$.

A program $P$ has a set of events, $events(P)$, such that for each
invocation event in $events(P)$, $events(P)$ also contains all possible corresponding response and
effect events. That is, each called operation can respond and take effect and the associated output is not under the program's control.

\[\all op:Op; in:Val\dot\\
\t1 inv(op,in)\mem events(P) \implies \all out:Val\dot \{res(op,out),\obs(op,out)\}\subseteq events(P) & (\refstepcounter{axiom}\arabic{axiom})
\label{events}
\]

\subsection{Traces}

The semantics of a program $P$ is described as a set of {\em finite\/} sequences of
events, referred to as {\em traces\/}.\footnote{Since we are interested in
defining a notion of correctness that readily relates to linearizability, 
and hence only safety properties \cite{GotsmanYang2011,Smith2017}, we do not
consider infinite sequences of events in our semantics.} 
In the following $t_i$ denotes the $i$th element of a trace $t$, and $\#t$ its length.

For each trace
$t$, each event is unique (similar events, e.g., calls to the same
operation, may be annotated by their relative position in the trace).

\[Trace \sdef \M \{t:\seq Event| \all i,j \leq \# t\dot i\neq j \implies t_i\neq t_j\}\O\]


The events of a trace and the order on these events are defined as
follows. 
\begin{align*}
events(t) &\sdef \{a:Event| \exi i\leq \#t\dot t_i=a\}\\
\order{t} &\sdef \{(a,b):Event\cross Event | \exi i,j\leq \# t\dot i < j \land t_i=a \land t_j=b\}
\end{align*}
Note that the order \!\!$\order{t}$ is a total order over the
events in $t$, as a trace describes exactly one execution of a program.

As a well-formedness condition on traces we postulate that an
invocation of an operation always occurs before both the associated
response and the associated effect. 
Similarly, a program step always occurs before its
effect. Also, the output value of an operation's effect is the same as that of the corresponding response event. For any trace $t$ we have

\[\M
(\all a:Op; out:Val\dot (\all j\leq \#t\dot t_j\mem \{res(a,out),eff(a,out)\} \implies \exi in:Val; i < j \dot t_i=inv(a,in))) \land\\
(\all s:T; p:PS\dot(\all j\leq\# t\dot t_j=\obs(s,p) \implies \exi i < j \dot t_i=step(s,p)))\land\\
(\all a:Op; out_r,out_e:Val\dot \{res(a,out_r),eff(a,out_e)\} \subseteq events(t) \implies out_r=out_e)\O
& (\refstepcounter{axiom}\label{wellformed}\arabic{axiom})
\]
%
Since responses remain in the order they appear in the program text, the constraint on the outputs of effects and responses prevents out-of-thin-air results.

\subsection{Programs}
\label{sec:porder}

The semantics of program $P$ on memory model $M$ is defined as
the set of traces using only events from $P$ and whose orders adhere to
the constraints prescribed by the partial order
$\!\!\Porder\!$. That is, if a pair of events $(a,b)$ is in $\!\!\Porder\!\!$ and $b$ is in an execution of $P$ on $M$ then $a$ occurs earlier in that execution. 
Hence the semantics of $P$ on $M$ is defined as

\[\semM{P} \sdef \{t:Trace|events(t) \subseteq events(P) \land \Porder \Subset \order{t}\}\]
where $\Porder \Subset \order{t}$ specifies whether an order is \emph{allowed} by $P$ on $M$,
formally defined as  

\[\Porder \Subset \order{t}\ \sdef~ \all (a,b):\Porder\dot b\mem events(t) \implies (a,b) \mem \order{t}\]
That is, for any event $b$ that occurs in trace $t$, if this event is
constrained to come after another event $a$ by $\!\!\Porder\!$, then
event $a$ must also occur in $t$ before event $b$. Note that
it is not suitable to use the simple subset relation here, i.e.,
$\Porder \subseteq \order{t}$, since trace $t$ will not, in general,
include all events $b$ that are constrained by $\!\!\Porder\!$.

Since we are aiming at the most general description of program semantics
under any memory model, we do not explicitly prescribe
$\!\!\Porder\!\!$. Instead, based on the understanding of concurrent
programs, concurrent objects and their interplay (as laid out in Section~\ref{sec:prog}),  we formulate certain characteristics of
$\!\!\Porder\!\!$ that are shared by all memory models, and provide a set of axioms 
in the remainder of this section.

\begin{figure}
 \centering
  \scalebox{0.9}{
  \includegraphics[viewport=7 760 388 825]{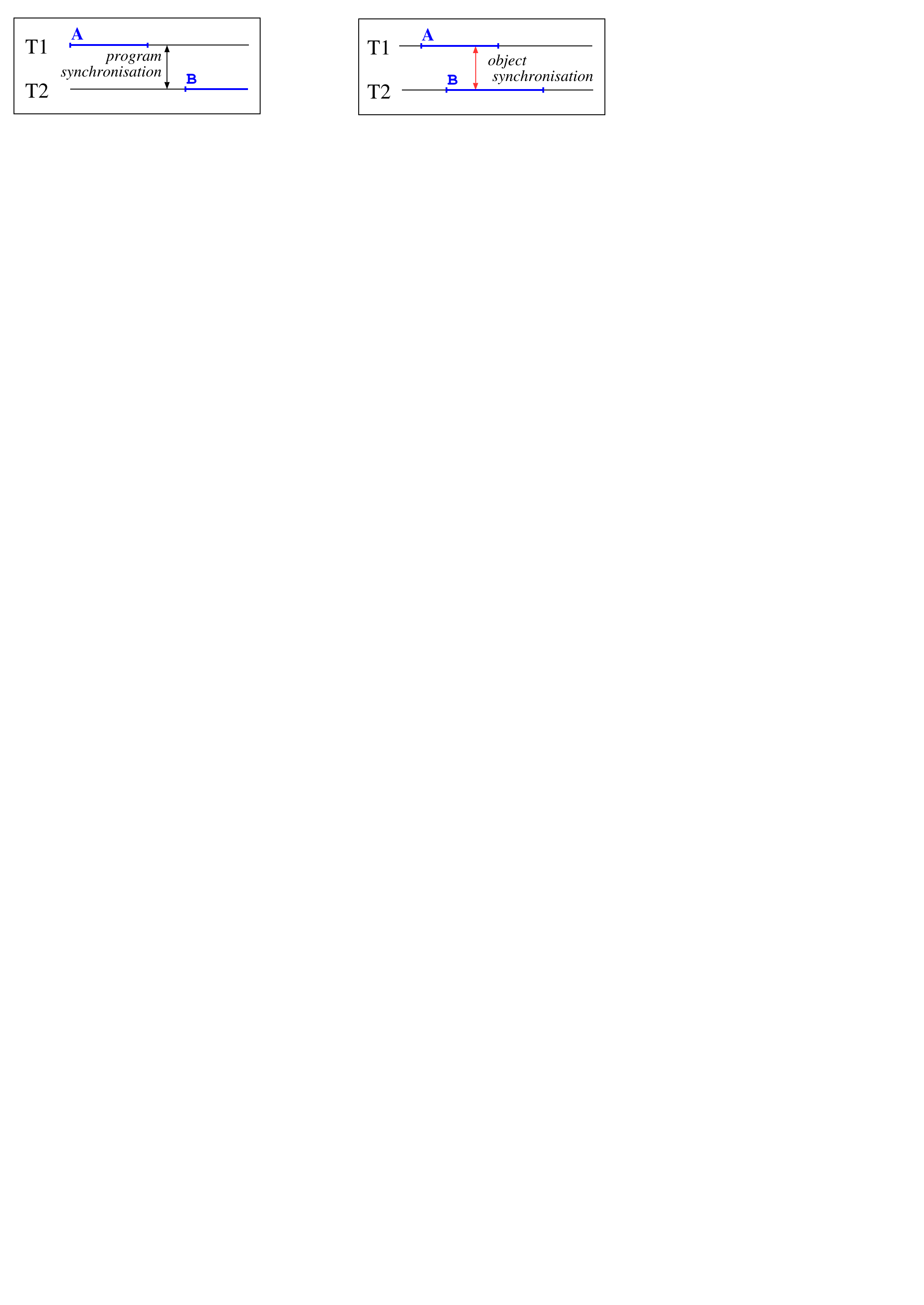}}
 \caption{Scenarios of synchronisations between threads}
 \label{fig:sync}
 \end{figure}

In the context of this work, we are interested in both the order of operation events and the order between operation
and program events in $\!\!\Porder\!\!$.  We define \emph{synchronisation}
(between threads) as an event on one thread affecting the occurrence of 
another event on another thread.
Synchronisation  requires a writing and a reading access to a shared variable 
(e.g., through an {\sf await} statement or a conditional).
We make the following assumptions on objects and clients:
\begin{itemize}
\item The state space of an object is encapsulated and hence the client does not share any 
  variables directly with the object; communication occurs only through input and output values of the object's
  operations.
\item Consequently, synchronisation between two threads can only occur between
  program events (referred to as \emph{program synchronisation})
  or between operation events  (referred to as \emph{object synchronisation}). The client cannot directly
  synchronise with a step in an operation that occurs between its invocation and
  response, but only with the invocation and response events themselves which serve as the interface
  between client and object. 
Figure~\ref{fig:sync} depicts the possible synchronisations by means of two scenarios:
The scenario on the left illustrates program synchronisation between the threads {\sf T1} and {\sf T2},
in which the program synchronises on two program steps which follow operation {\sf A} and precede
operation {\sf B}, respectively. An example of this type of synchronisation is given by the instructions {\sf z=1} and {\sf await(z=1)} in Figure~\ref{fig:client}. The scenario on the right illustrates operations
{\sf A} and {\sf B} overlapping and  synchronising on the object level,
outside the control of $P$. 
\item A client cannot enforce a flush within an operation; 
  the operation's implementation is outside the client's control.
\end{itemize}

From these assumptions we can deduce that 
$\!\!\Porder\!\!$ can only enforce the invocation of an operation to occur
\emph{before} another event, which is not the effect of the operation\footnote{The other event can be the response of the operation since partial orders are reflexive and hence $(res(c),res(c))$ is in $\Porder$ for any $c$.}, if and only if the response is also enforced to occur before the event.
(The implication from right to left holds due to the well-formedness condition (\ref{wellformed}).)

\[\all e, inv(a,in), res(a,out):events(P) \dot \\
\t1 e\neq eff(a,out)\implies
    ((inv(a,in), e)\mem \Porder 
    \iff
    (res(a,out), e)\mem \Porder) & (\refstepcounter{axiom}\label{enf-inv-before}\arabic{axiom})
\]

Similarly, $\!\!\Porder\!\!$ cannot enforce the response of an
operation to come \emph{after} another event, which is not the effect of the operation, unless it also enforces
the invocation to come after the event.  (As above, the implication
from right to left holds due to well-formedness (\ref{wellformed}).
  )

\[\all e, inv(a,in), res(a,out):events(P)\dot \\
\t1 e\neq eff(a,out)\implies
    ((e, res(a,out))\mem \Porder 
    \iff
    (e,inv(a,in))\mem \Porder) & (\refstepcounter{axiom}\label{enf-res-after}\arabic{axiom})
\]

A similar condition holds for the order of effect events with respect to
program events. Due to an object's implementation, the effect of an
operation might occur before its response event (e.g., if the operation
is implemented with fence instructions which prevent a delay in its
effect), which is outside the control of $P$.
Hence
$\!\!\Porder\!\!$ cannot enforce the effect of an operation to come
\emph{after} another \emph{program} event unless it also enforces the invocation
to come after the program event.
(Again, the implication from right to left holds due to well-formedness (\ref{wellformed}).)

\[\all e, inv(a,in), eff(a,out):events(P)\dot \\
    \t1 e \mem \PEv\implies
    ((e, \obs(a,out))\mem \Porder 
    \iff
    (e,inv(a,in))\mem \Porder) & (\refstepcounter{axiom}\label{enf-obs-after-ps}\arabic{axiom})
\]

In some memory models, the order in which two effects can occur can be constrained 
(e.g., in TSO the order of observable effects follows the order of the observed events), but only
if the order of the corresponding events is also enforced. Any order that is caused by the nature
of a particular implementation (e.g., an additional fence instruction in the code) is outside
the control of $P$ (similarly to the case above).
This means, if the effects of two operations are ordered 
then the order of the operations must be enforced, 
and hence the response of one operation must occur before the 
invocation of the other.

\[\all inv(a,in), \obs(a,out), res(c), \obs(c):events(P)\dot \\
    \t1 ((\obs(c), \obs(a,out))\mem \Porder \land c\neq (a,out))
    \implies 
    (res(c),inv(a,in))\mem \Porder 
\]

A similar condition holds if $\!\!\Porder\!\!$ enforces an operation effect to occur after
the invocation, response or effect event of another operation. This is only possible if the two
operations are ordered (and hence the response of the first must occur before
the invocation of the other). If the operations are not ordered by $\!\!\Porder\!$, they may occur in
either order (and hence the effects may occur in either order) 
or they may overlap. For overlapping operations,
$\!\!\Porder\!\!$ cannot enforce an order of object events since any object synchronisation is beyond the control
of $P$. Therefore we can generalise the above axiom as follows.

\[
  \all inv(a,in_a), res(a,out_a), \obs(a,out_a), inv(b,in_b), \obs(b,out_b):events(P)\dot\\
  \t1 \all e : \{inv(a,in_a), res(a,out_a), \obs(a,out_a)\}\dot\\
  \t2  ((e, \obs(b,out_b) \in \Porder \land a \neq b)
  \implies
  (res(a,out_a), inv(b,in_b)) \in \Porder & (\refstepcounter{axiom}\label{enf-order-obs}\arabic{axiom})
\]

As a consequence of these observations we deduce that 
an order between object events of two operations can only be enforced by $\!\!\Porder\!\!$ if 
$\!\!\Porder\!\!$ also enforces that these two operations do not overlap. 

\begin{lemma}[Enforced ordering on object events]\label{porderLemma}
\[\all inv(a,in_a), res(a,out_a), \obs(a,out_a), inv(b,in_b), res(b,out_b), \obs(b,out_b):events(P)\dot\\
\t1 \all e : \{inv(a,in_a), res(a,out_a), \obs(a,out_a)\}; 
        e' : \{inv(b,in_b), res(b,out_b), \obs(b,out_b)\} \dot \\
    \t2       ((e, e') \in \Porder  \land a \neq b)    ~\implies~ (res(a,out_a), inv(b,in_b)) \in \Porder
\]
\end{lemma}

The proof follows the simple application of Axioms  (\ref{enf-inv-before}) to (\ref{enf-order-obs})
 to all combinations of invocation, response and effect events of the two operations.

\section{Semantics of objects under weak memory models}
\label{sec:objectsemantics}

To define object refinement, our notion of correctness, in Section~\ref{sec:wmmref}, we need to constrain the behaviour of a program to a particular
object or collection of objects (called an \textit{object system} in \cite{Filipovic-LinvsRef2010}). In either case, the interface between the program and the one or more objects is the operations of those objects, the outputs of which, when the operations are called in a given sequence, reflect the objects'  behaviour including any interactions between them. For ease of presentation, we consider a single object only, but the results also hold for a collection of interacting objects whose interface is the operations of all objects in the collection. 

Let $\resIR{t}$ denotes the trace $t$ restricted to invocation and response events. The semantics of an object is given as a set of \emph{histories}, where
each history is a trace with only invocation and response events.

\[History \sdef \{t:Trace|\resIR{t}=t\}\]

\subsection{Object implementation under weak memory models}
\label{subsec:semC}

An object implementation $C$ has a set of object events, $events(C)$, 
and, on a particular memory model $M$, a prefix-closed set of histories 
made up of those events, $\semM{C}$.\footnote{The set of histories of $C$ would be derived in an operational manner based on the weak memory model semantics. This may involve two passes in which firstly effects and their placement are taken into account, and secondly response values are added to reflect this placement, and effects are discarded.}

For any object implementation $C$, $P[C]$ denotes the object $C$ operating in program $P$. 
It is only defined when
all invocation events of $P$ are events of $C$. 
The semantics of $C$ operating in $P$ on memory model $M$, $\semM{P[C]}$, is given as those traces 
of $P$ on $M$ whose object events correspond to a history of $C$ on $M$.

\[\semM{P[C]} \sdef \{t:\semM{P}|\exi h:\semM{C}\dot \resIR{t}=h\}\]

\subsection{Object specification under weak memory models}
\label{subsec:semA}

An object specification $A$ similarly has a set of invocation and response events,
$events(A)$, and a prefix-closed set of histories, $\sem{A}$ (where prefixes are restricted
to complete histories \cite{HeWi90} in which the final event cannot be an invocation). Since
$A$ represents a typical specification found in a software library,
its set of histories is \emph{independent} of the memory model (hence there is no subscript).  
Any weak memory model behaviour is absent from its histories due to
its operations being {\em atomic\/}, i.e., they respond and take effect immediately 
and hence occur without interference from other operations.
That is, every history in $\sem{A}$ is a sequence of invocation and response pairs, 
i.e., every invocation is followed by its response, and no overlapping of operations is possible.

\[ \hspace*{-0.4cm}   
    \all h : \sem{A} \dot \\
  \hspace*{-0cm}   
         (\all i \leq \# h\dot h_i=inv(a,in) \implies (i<\#h  \land \exi out:Val \dot h_{i+1}=res(a,out))) \land \\
         (\all i \leq \# h\dot h_i=res(a,out) \implies (i\neq 1 \land \exi in:Val \dot h_{i-1}=inv(a,in)))
  ~& (\refstepcounter{axiom}\label{seqSpec}\arabic{axiom})
\]



\begin{figure}
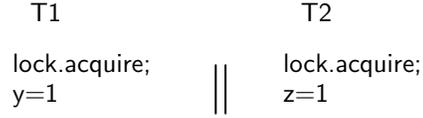

  \vspace{-8ex}
\[
 \qquad \qquad \qquad \quad \qquad\qquad\parbox[t]{2.6cm}{~~{\sf T1} \vspace*{2ex}\\ 
  {\sf lock.acquire;\\ y=1}} 
     \parbox[h]{1cm}{\vspace*{2cm}{\resizebox{!}{0.5cm}{$\parallel$}}} 
    \parbox[t]{3cm}{~~{\sf T2}\vspace*{2ex}\\ {\sf lock.acquire;\\ z=1}}
\]
\caption{Client program using lock}
 \label{fig:lock}
\end{figure}

To capture atomicity of specifications in our semantics, the behaviour of a client using specification 
$A$ is restricted
to those traces where only one operation is active at a time.
For example, suppose the specification of a lock object, {\sf lock},
has an operation {\sf acquire} which waits until the value of a
variable of the object, {\sf x}, is 1 and sets it to 0, i.e., {\sf
  acquire} is specified as {\sf await(x=1); x=0}. Assuming {\sf x} is
initially 1, in the program of Figure~\ref{fig:lock} the intention would
be that only one of {\sf y} or {\sf z} would be set to 1.

On SC, this intention is achieved when the invocation of {\sf acquire}
which occurs second does not happen until after the first {\sf acquire} has taken effect.
On TSO and assuming T1 and T2 are
running on different cores, the intention is achieved when the
invocation of {\sf acquire} which occurs second does not happen until
after the flush of {\sf x} from the {\sf acquire} which occurs
first. This flush might be delayed. On ARM and POWER (again assuming T1 and T2 are on different cores), the intention is achieved when the second invocation of {\sf acquire} does not happen until after the write to {\sf x} by the first invocation has been seen by all threads (see description of non-multi-copy atomicity in Section~\ref{sec:observations}). 

In all cases, the intention is met when the second occurrence
of {\sf acquire} is not invoked before the effect event of the
first occurrence. In general, for any object specification to behave
as intended within a client context, an operation invocation does not occur
before the effect event of a previously invoked operation. (Note that due to Axiom~(\ref{seqSpec}) invocations also must occur after responses of a previous operation.)

Provided all invocation events of a program $P$ are events of a
specification $A$, $P[A]$ denotes the program $P$ operating with an abstract
object whose behaviour satisfies~$A$.
To model atomicity, operations can only be invoked after previous operations have taken effect (as motivated above).
The semantics of $P[A]$ is given as follows.
\[\semM{P[A]} \sdef \{~t:\semM{P}|\M  \exi h:\sem{A}\dot \resIR{t}=h ~\land\\
                                   \all c:Op\cross Val; k\leq \#t \dot t_k=inv(c) \implies\\
                                   \t1 \all a:Op;in:Val; i < k \dot t_i=inv(a,in)\implies \\
                                    \t3 \exi out:Val; j < k \dot t_j=\obs(a,out)  \O ~                        
\}\]



\section{Object refinement}
\label{sec:wmmref}

Correctness of an object is defined from the client program's point of
view. Such a program can only observe changes to {\em program
  variables\/}, i.e., variables that are not defined locally on a
thread or as part on an object.  Let $\obsT{t}$ denote the {\em
  observable behaviour\/} of a trace $t$, i.e., the sequence of
effect events of program steps which write to global program
variables. If for two traces $t$ and $t'$ we have $\obsT{t}=\obsT{t'}$, we call them \emph{matching} traces.

A program
$P$ using $C$ on memory model $M$ refines $P$ using
$A$ on $M$ when any observable behaviour of the former is a possible observable behaviour of the latter, i.e., each concrete trace has a matching abstract trace. We refer to this property as {\em \wmmRef\/}$\!\!$.

\begin{definition}{\WMMRef}
\label{wmmRefdef}

\[P[A] \refsto_M P[C] ~\sdef~ \all t:\semM{P[C]}\dot \exi t':\semM{P[A]}\dot \obsT{t'}=\obsT{t}\]
\end{definition}

An object implementation $C$ refines an object specification $A$ under weak memory model $M$
if for all possible client programs $P$, $P$ using $C$ weak-memory trace refines $P$ using $A$ under memory model $M$. We refer to this property as \emph{object refinement}.

\begin{definition}{Object Refinement under memory model $M$}
\[ A \refsto_M C ~\sdef~ \forall P \dot P[A] \refsto_M P[C]\]
\end{definition}

If $A \refsto_M C$ we say that $C$ is correct with respect to $A$ under memory model $M$.

\section{Linearizability}
\label{sec:lin}


Linearizability relates histories of an object implementation, which may have
{\em pending\/} invocations, i.e., invocations for which there is no response, to histories of an object specification which do not
\cite{HeWi90}. To do this, it needs to {\em complete\/} the implementation
histories. This can be done by adding a response
when a pending invocation is deemed to have taken effect, and removing the invocation when it has not
\cite{HeWi90}. 
For example, consider a history comprising a read operation of a
variable $x$ occurring on a thread T1 concurrently with a write operation to $x$
on thread T2 where the latter has not yet responded. If the read operation
returns the value from the write operation, we can assume the write operation
has taken effect and hence we add a response to the history. If the read
operation returns the value of $x$ from before the write operation, we can
assume the latter has not taken effect and remove its pending invocation.\footnote{In this case, we could also add a response since the write could have taken effect, but after the read took effect. However, to be consistent with the original definition of linearizability, we allow invocations to be removed.}

To define linearizability, functions are put in place for adding
responses, and removing invocations from histories. The function $ext$ returns
the set of traces which extend a given trace with a sequence of responses
such that the result is still a trace, i.e., responses are only added for
pending invocations.

 \[ext(t)\sdef \{t\cat tr:Trace | \all i \leq \#tr \dot \exi c: Op\cross Val \dot  tr_i=res(c) \}
\]

The function $comp$ returns the trace resulting from the removal of all invocations from a given trace which have neither an effect nor a response.

\[comp(t) \sdef   \left
\{ \begin{array}{ll}
\!\!\emptyseq \mbox{, ~if  $t=\emptyseq$} \\
\!\!comp(\tail t)\mbox{, ~if  $NoResp(t)$}\\
\!\!\lseq\head t\rseq \cat comp(\tail t)\mbox{, ~otherwise}
\end{array}
\right.
\]
where $NoResp(t) \sdef \exi a:Op; in:Val\dot$\\
 $~~~~\t5 head(t)=inv(a,in) \land \nexi out:Val; i\leq \# t\dot t_i\mem \{res(a,out), eff(a,out)\}$\\

The following formalisation of the standard definition of linearizability is
based on that of Derrick et al.\ \cite{DSW10mvpolin} which has been proved
to correspond to the original definition by Herlihy and
Wing \cite{HeWi90}. (Note that $comp(h^+)$ removes all pending invocations from $h^+$ since histories do not have effect events.)

\begin{definition}{Linearizability (standard definition)}

\[C \linM A ~\sdef~ \M \all h:\semM{C} \dot \exi h':\sem{A} \dot \exi h^+:ext(h)\dot  comp(h^+) \sim h' \land \ \supclass_{comp(h^+)}\, \subseteq \ \supclass_{h'}\O\]
\end{definition}
where $t \sim t'$ denotes that $t$ and $t'$ are {\em thread equivalent\/}, i.e.,
when restricted to the events of any one thread they have the same sequence of
invocations and responses, and $\supclass_{t} \sdef \{(res(c),inv(d)):\order{t} \}$,
\label{supclass_def}
i.e., $\supclass_{t}$ captures the order between operations in a trace (where an operation comes before another if its response is before the other's invocation).

The intuition behind the definition is that operations which are overlapping in
$comp(h^+)$ are not ordered by $\supclass_{comp(h^+)}$ and, with $\supclass_{h'}$ being a superset of $\supclass_{comp(h^+)}$,
can occur in any order in $h'$.
For example, the overlapping
operations $B$ and $C$ of the implementation history of Figure~\ref{fig:lin} can
occur in any order in a linearization of that history. This is equivalent to
letting the linearization points of $B$ and $C$ occur anytime between the
respective operations' invocations and responses.

Importantly, the definition is {\em compositional\/} (this property is proved for the above definition by Herlihy and Wing who refer to it as {\em locality\/} 
 \cite{HeWi90}). In the case when the object
implementation $C$ is a collection of interacting objects, compositionality
allows us to prove that $C$ is linearizable to a specification of a similar
collection of interacting objects by proving each individual object
implementation is linearizable to the corresponding object specification.

\vspace*{2ex}

 In the remainder of this section, we prove that standard linearizability is sound and complete with respect to our definition of object refinement. To facilitate the proofs, we introduce the following lemmas on completions of traces
whose proofs are included in Appendix~\ref{app:lemmas}.






%

\begin{lemma}\label{comp} If the events of a trace $t$ are events of a program $P$ then so are the events of any completion of~$t$.
\vspace*{-1ex}
\[\all P\dot\all t:Trace\dot\all t^+:ext(t)\dot
 events(t)\subseteq events(P) \,\implies\, events(comp(t^+)) \subseteq events(P)\]
\end{lemma}

\begin{lemma}\label{Subsetcomp} If a trace $t$ is allowed by a program $P$ on memory model $M$ then so is any completion of trace $t$ that only adds responses for operations whose effects have occurred.

\[\all P, M\dot\all t:Trace\dot \all t\cat tr:ext(t)\dot \\
~~~~~ (\all i\leq \# tr; c:Op \cross Val \dot tr_i=res(c) \implies \exi j\leq \# t\dot t_j=eff(c)) \land \Porder\Subset \order{t} \implies \Porder\Subset \order{comp(t\cat tr)}\]
\end{lemma}

\begin{lemma}\label{subset}
The operation order of a completion of a trace $t$ is a subset of that of $t$.
 \[\all t:Trace\dot \all t^+:ext(t)\dot \ \supclass_{comp(t^+)}\ \subseteq\ \supclass_{t}\]
 \end{lemma}


\subsection{Soundness}\label{sec:soundness}

The soundness of standard linearizability in the context of weak memory model $M$ 
can now be proved in relation to our notion of object refinement under $M$.

\begin{theorem} \label{soundness} If an object implementation $C$ linearizes
  with an object specification $A$ on memory model $M$ then for all client programs
  $P$, $P[C]$ is a weak-memory trace refinement of $P[A]$ on $M$.

\[C \linM A~ \implies ~\all P\dot P[A] \refsto_M P[C]\]
\end{theorem}

\newenvironment{s-equation}{%
\refstepcounter{soundprops}
\renewcommand\theequation{S\thesoundprops}
\begin{equation}}
{\end{equation}}
\vspace*{2ex}

The proof of this theorem takes any concrete history $h$ of $C$ and any abstract history $h'$ of $A$, where the latter is a linearization of the former, 
and any trace $t$ of a program calling $C$ that corresponds to the concrete history $h$.
From these ingredients we construct an abstract trace $t'$ such that $t'$ corresponds to the linearization $h'$. 
That is, we move the object events of $t$ such that the operations do not overlap and also occur in the same order
as in $h'$ in the resulting trace $t'$.
It then remains to show that 
the trace that is constructed in this way is indeed a trace of the client calling the abstract object 
(i.e., that $t'$ is a trace of $P[A]$).

\vspace*{2ex}

\proof ~~
Expanding $\linM\!\!$ and $\refsto_M$ we have
\[
\hspace*{-3ex}
(\all h:\semM{C} \dot \exi h':\sem{A} \dot \exi h^+:ext(h)\dot
comp(h^+)\sim h' \land\ \supclass_{comp(h^+)}\ \subseteq\ \supclass_{h'})\implies\\
\t1 \all P\dot \all t:\semM{P[C]}\dot \exi t':\semM{P[A]}\dot \obsT{t'}=\obsT{t}\]
We need to prove that the consequent holds whenever the
antecedent does. For any $P$, either $\semM{P[C]}=\emptyset$ and the consequent
trivially holds, or there exists a trace $t$ in $\semM{P[C]}$. In the latter
case, we have that there exists an $h$ in $\semM{C}$ such that $\resIR{t}=h$
(from the definition of $\semM{P[C]}$).\vspace*{1ex}

\noindent
Given such a $t$ and $h$, when the antecedent holds we also have an $h'\in \sem{A}$ and an $h^+ \in ext(h)$ 
such that 
\begin{s-equation}
  comp(h^+)\sim h' ~\land~  \supclass_{comp(h^+)}\ \subseteq\ \supclass_{h'} \label{sp_1}
\end{s-equation}
\noindent \!\!There may be a number of choices for $h^+$ based on how many pending invocations are given responses. We choose $h^+$ so that there is an added response for exactly those pending invocations whose effects occur in $t$. This is always possible since we know that there exists at least one extension and related abstract history. Call them $h^+_0$ and $h'_0$, respectively. $h^+_0$ cannot have less than the required responses. If it did, $comp(h^+)$ would be left with a pending invocation (with an effect) but no response. Hence, $comp(h^+)\sim h'$ would not hold. If $h^+_0$ has more than the required responses, since $\sem{A}$ is prefixed-closed, we can find an $h'$ which is a subsequence of $h'_0$ which does not have the additional operations corresponding to the extra responses.\footnote{Since there is at most one pending invocation per thread, such an $h'$ will be in the prefix-closed set $\sem{A}$.}  This $h'$ will satisfy (\ref{sp_1}) for our chosen $h+$.

Since $t \in \semM{P[C]}$, we can deduce from the definition of $\semM{P[C]}$ that $t\mem \semM{P}$,
and with the definition of $\semM{P}$, we have that
\begin{s-equation}
 events(t)\subseteq events(P) ~\land\  \Porder \Subset \order{t}  \label{sp_2}
\end{s-equation}
%
\!\!Let $t^+ \mem ext(t)$ be the trace that extends $t$ with the same sequence of responses that $h^+$ extends $h$. 
It follows that 
\begin{s-equation}
\resIR{t^+}=h^+  ~\land ~ \resIR{comp(t^+)} = comp(h^+)  \label{sp_2a}
\end{s-equation}
%
%
\!\!From (\ref{sp_2}), Lemma~\ref{comp} and Lemma~\ref{Subsetcomp} (and since we choose $h^+$ to only add responses to $h$ for operations whose effects occur in $t$) we have 
\begin{s-equation}
 events(comp(t^+)) \subseteq events (P) ~\land~  \Porder \Subset \order{comp(t^+)} \label{sp_3}
\end{s-equation}
Since  $\resIR{t^+}=h^+$ (\ref{sp_2a}) and $comp(h^+)\sim h'$ (\ref{sp_1}) we can deduce that 
\begin{s-equation}
 comp(\resIR{t^+}) \sim h' \label{sp_4}
\end{s-equation}

Central to our proof is the fact that we can  construct the abstract trace $t'$ appearing in the consequent of the theorem from trace $comp(t^+)$ and history $h'$. This is done 
by means of a transformation function $trans$,
which reorders object events in $comp(t^+)$ to match the order in $h'$, while maintaining the order of program events.
We let $t'=trans(comp(t^+),h',\emptyseq)$ using the following definition of the transformation function.
\begin{definition}[Transformation function]\label{transdef}
  Let $a\mem Op$ be an operation, $in, out\mem Val$ be inputs and outputs, 
and $r \mem \seq \ObjEv$ a (remainder) sequence that temporarily stores invocations which will be placed 
into the resulting trace at a later point. The latter allows the invocations in the original trace to be reordered in the resulting trace.\\
Then  $trans(t,h',r)$ is equal to either\\
$
\begin{array}{rll}
(T1) & \multicolumn{2}{l}{\lseq inv(a,in), eff(a,out),res(a,out)\rseq\cat trans(\tail t,\tail(\tail h'),r)}\\
 &  &\mbox{if } \head t = \head h' = inv(a,in) \land  \exi i\leq \#t \dot t_i=res(a,out), \\
(T2) & trans(\tail t,h',r\cat \lseq inv(a,in)\rseq) & \mbox{if } \head t = inv(a,in) \land \head h' \neq inv(a,in),\\
(T3) & trans(\tail t, h', r) & \mbox{if } \head t \mem \{res(a,out), \obs(a,out)\},\\
(T4) & \lseq \head t\rseq \cat trans(\tail t,h', r) & \mbox{if } \head t \mem \PEv \land \nexi\, i\leq \#r\dot  r_i = \head h',
\ \mbox{or}\\
(T5) &   \multicolumn{2}{l}{\lseq inv(a,in), eff(a,out),res(a,out)\rseq\cat trans(t,\tail(\tail h'),r - \{inv(a,in)\})}\\
 &  & \mbox{if } \exi i\!\leq\! \#r\dot r_i = \head h' = inv(a,in) \wedge \exi i\!\leq\! \#t\dot t_i=res(a,out).
\end{array}
$
\end{definition}

Note that (T1) -- (T5) cover all cases when $t$ has a response for each invocation (as $comp(t^+)$ does). If the head of $t$ is an invocation, it is either added to $t'$ (along with the corresponding effect and a matching response) (T1) or added to $r$ (T2). If it is a response or effect, it is removed (T3). If it is a program event, and the head of $h'$ is not in $r$, it is added to $t'$ (T4). If the head of $h'$ is in $r$, it is added to $t'$ (along with the corresponding effect and a matching response) (T5). Since $comp(\resIR{t^+}) \sim h'$ (\ref{sp_4}), all invocations added to $r$ will eventually be processed via the recursive definition.  

It is easy to prove that cases (T1) -- (T5) of Definition~\ref{transdef} ensure that the newly constructed trace $t'$ contains the same invocation and response events as $comp(t^+)$ and
its restriction to invocation and response events matches $h'$.
\begin{s-equation}
  events(\resIR{t'})=events(\resIR{comp(t^+)}) ~~\land~~ \resIR{t'}=h' 
 \label{sp_5}
\end{s-equation}
\!\!Cases (T1) and (T5) in particular ensure that in the transformed trace the operations do not overlap (as invocations, effects and responses of each operation are added to $t'$ in one step).
\begin{s-equation}
\begin{array}{l}
\all c:Op\cross Val; k \leq \#t'\dot t'_k=inv(c) \implies\\
                              \t1     \all a:Op;in:Val; i < k \dot t'_i=inv(a,in)\implies 
                                    \exi out:Val; j < k\dot t'_j=\obs(a,out)  
\end{array}
 \label{sp_6}
\end{s-equation}

With case (T4) we also ensure that program steps are in the same order in $t$ and $t'$. 
\begin{s-equation}
 \resP{t'}=\resP{t} \label{sp_6a}
\end{s-equation}
where $\resP{t}$ denotes the trace $t$ restricted to program events. Hence, we have
\begin{s-equation}
 \obsT{t'}=\obsT{t} \label{sp_7}
\end{s-equation}

To prove the consequent of Theorem~\ref{soundness} it remains to show that
$t' \in \semM{P[A]}$.
Given the definition of $\semM{P[A]}$ and (\ref{sp_5}) and (\ref{sp_6}) this reduces to proving that
$t' \in \semM{P}$, i.e., $t' \mem Trace$, $events(t') \subseteq events(P)$ and $\Porder \Subset \order{t'}$.

\begin{itemize}
\item[i.)]  $t' \mem Trace$: \\
         With $t \in \semM{P[C]}$ and Lemmas~\ref{comp} and \ref{Subsetcomp}
         it follows that $comp(t^+) \in \semM{P}$ and hence is a $Trace$. \\
         With $events(\resIR{t'})=events(\resIR{comp(t^+)})$ (\ref{sp_5}) and $\resP{t'}=\resP{t}$ (\ref{sp_6a}) and the fact that effects are only added to $t'$ to match existing responses ((T1) and (T5)), it follows that
         all events in $t'$ are unique.\\
         Due to cases (T1) and (T5) it is guaranteed that the order of 
         object events in $t'$  is well-formed, i.e., for all $inv(a,in)$ in $t'$ we have
         $inv(a,in) \order{t'} res(a,out) \land inv(a,in) \order{t'} \obs(a,out)$.\\
         From case (T4) it follows that the order of
         program events (program steps and their effects) is maintained in $t'$, i.e., for all $step(s,p)$ and $\obs(s,p)$ in $t'$ we have 
         $step(s,p) \order{t'} \obs(s,p)$.\\
         From the above it can be deduced that since $t$ is a well-formed trace, $t'$ is also 
         a well-formed trace, i.e., 
         \begin{s-equation}
             t' \mem Trace  \label{sp_8} 
         \end{s-equation}
\item[ii.)] $events(t') \subseteq events(P)$: \\
         With $events(comp(t^+)) \subseteq events(P)$ ((\ref{sp_2}) and Axiom (\ref{events}))
         and $events(\resIR{t'})=events(\resIR{comp(t^+)})$ (\ref{sp_5}) and $\resP{t'}=\resP{t}$ (\ref{sp_6a}) and the fact that effects are only added to $t'$ to match existing responses ((T1) and (T5)) and these are in $events(P)$ due to Axiom~(\ref{events}), 
         it follows that
         \begin{s-equation} events(t') \subseteq events(P)\label{sp_9}
         \end{s-equation}
\item[iii.)] $\Porder \Subset \order{t'}$:\\
         To prove this property it must be ensured that
         $\all (a,b):\Porder\dot b\mem events(t') \implies (a,b) \mem \order{t'}$. 
         This property is split into the following four cases.
         \begin{itemize}
            \item[a.)] The order between two program events that is enforced by $\Porder$ is maintained in $t'$.
                       \[
                         \all a,b:\PEv\dot (a, b)\in\Porder \land b\in events(t')\implies 
                                    (a,b)\in\order{t'}\]
            \item[b.)] If $\!\!\Porder\!\!$ enforces that a program event has to occur before an object event
                       then this order is maintained by $t'$.
                       \[\all p :\PEv; e\in\ObjEv\dot (p,e)\in\Porder \land e\in events(t')\implies (p,e)\in\order{t'} \]
            \item[c.)] If $\!\!\Porder\!\!$ enforces that an object event has to occur before a program event
                       then this order is maintained by $t'$.
              \[
                   \all p :\PEv; e\in\ObjEv\dot (e, p)\in\Porder \land p\in events(t')\implies (e, p)\in\order{t'}\]
            \item[d.)] The order between two object events that is enforced by $\Porder$ is maintained in $t'$.
              \[
                    \all a,b:\ObjEv\dot (a, b)\in\Porder \land b\in events(t')\implies (a,b)\in\order{t'}\]
         \end{itemize}
         
\end{itemize}

These four remaining properties on $t'$ rely firstly on 
assumptions on $\!\!\Porder\!\!$ relating to constraints on object events that might be 
enforced by $P$ and $M$, which were outlined in Section~\ref{sec:progsemantics},
and secondly on properties of $trans$ relating to the order of program events in $t'$ which can be 
deduced from Definition~\ref{transdef}. The proofs are provided in Appendix~\ref{app:sound}. \hfill $\Box$

\newenvironment{c-equation}{%
\refstepcounter{compprops}
\renewcommand\theequation{C\thecompprops}
\begin{equation}}
{\end{equation}}

\newenvironment{c-align}{%
\refstepcounter{compprops}
\renewcommand\theequation{C\thecompprops}
\begin{align}}
{\end{align}}

\subsection{Completeness}\label{sec:completeness}

\begin{theorem} If we have an object implementation $C$ and specification $A$ such that, for all programs $P$, $P[C]$ is a weak-memory trace refinement of $P[A]$ on memory model $M$ then $C$ linearizes to $A$ on $M$.
\[(\all P\dot P[A] \refsto_M P[C])~ \implies ~C \linM A\]
\label{compTheorem}
\end{theorem}

%
%
%
The proof of this theorem is based on the notion of a ``strongest'' client program $\widehat{P}$ for each implementation history $h$ (the details of which are explained below). When such a strongest client program exists, the antecedent of the theorem implies $\widehat{P}[A] \refsto_M \widehat{P}[C]$ and we use this fact to prove that there exists an $h'\mem \sem{A}$ such that $h$ linearizes to $h'$. In the case that such a strongest client program does not exist, we show that either (i) the theorem is trivially true, or (ii) there must be an $h'\mem \sem{A}$ such that $h$ linearizes to $h'$. Applying these arguments to all $h$ in $\semM{C}$ results in the theorem's consequent being true.\\

The strongest client program $\widehat{P}$ of a history $h$ is one which

\begin{enumerate}
\item allows $h$ to occur enforcing the order of its operations, and 
\item records every invocation and response in a global variable without delay. 
\end{enumerate}

Property~1 requires that $\widehat{P}$ enforces the order of operations (i.e., the response/invocation order) to be as in $h$. Although $\widehat{P}$ cannot control the outputs of operations, if it calls the operations with the inputs in $h$ the corresponding outputs will be possible (since $h$ is an implementation history). For operations on a particular thread, the required order is enforced by $\widehat{P}$'s program order. For operations on different threads, it can be achieved using inter-thread program synchronisation where necessary.
For example, if the implementation history $h$ refers to two threads, $n$ and $m$, and the response of $OpA(x)$ on $n$ occurs before the invocation of $OpB(y)$ on $m$ in $h$ then $\widehat{P}$ could be of the form \[...;u=OpA(x);z=1;... || ...; await(z=1);v=OpB(y);...\] where initially $z=0$ and the elided parts of the program do not change $z$.

Property~2 can be achieved,  for example, with $\widehat{P}$
  having global variables $var_{inv}$ and $var_{res}$ and every operation call $OpA(x)$
  on thread $n$ which returns a value appearing in $\widehat{P}$ as
  $var_{inv}=(n,OpA,x); fence; y=OpA(x); var_{res}=(n,OpA,y); fence$, and every operation call $OpA(x)$
  which doesn't return a value as $var_{inv}=(n,OpA,x); fence; OpA(x);
  var_{res}=(n,OpA,\bot); fence$.  
Note that the recording of invocations and responses using $var_{inv}$ and $var_{res}$ is decoupled from the actual operation invocations and responses, 
and hence interleaving of the recording events from different threads is possible.  

It is not necessarily the case that matching traces of such a strongest client program 
have the same sequence of invocations and responses.
This is due to the fact
that an invocation might be recorded (via $var_{inv}$) but the operation not (yet) called, 
or an invocation and response have occurred but the response is not yet recorded
(recall that traces are prefix-closed).
Nevertheless,  for each abstract trace that 
matches a concrete trace we can always find  another abstract trace that
also matches the concrete trace and additionally has the same sequence of invocations and responses 
as the completed concrete trace.

\begin{lemma}\label{matchingTraces}
 For an object implementation $C$ which is an object refinement of a specification $A$ under $M$, 
and any client program ${P}$ that records every invocation and response of an operation without delay, we have

\[\all t : \semM{{P}[C]} \dot (\exi t':\semM{{P}[A]}\dot \obsT{t}=\obsT{t'})\implies \\
    \t2 \exi t'':\semM{{P}[A]} \dot \obsT{t}=\obsT{t''}   \land \exi t^+:ext(t) \dot \resIR{comp(t^+)}\sim \resIR{t''}    \]
\end{lemma}

The proof of this lemma is based on the prefix closure of traces and 
can be found in Appendix~\ref{app:complete}.
\vspace*{2ex}

\begin{myproof}{of Theorem~\ref{compTheorem}}

Expanding $\linM\!\!$ and $\refsto_M$ we have

\[
(\all P\dot \all t:\semM{P[C]}\dot \exi t':\semM{P[A]}\dot \obsT{t'}=\obsT{t})\implies\\
\hspace*{2ex}
 \all h:\semM{C} \dot \exi h':\sem{A} \dot \exi h^+:ext(h)\dot
comp(h^+) \sim h' \land\ \supclass_{comp(h^+)}\ \subseteq\ \supclass_{h'}\]

Assuming the antecedent is true, we prove the consequent for any $h\mem \semM{C}$. It will not be possible to construct a strongest client program for $h$ when the synchronisation between two threads, added explicitly or via a fence, changes the behaviour of the operation occurring after the synchronisation. In the example above, if synchronising using $z$ causes $OpB(y)$ to result in a different output then the resulting program will not allow $h$ to occur. This situation would arise, for example, on TSO where the effect of $OpA(x)$ would necessarily occur before the effect of $z=1$ due to the FIFO order of the per-core buffer. In other words, there will be no strongest client program when the behaviour of one operation ($OpB(y)$ in the example) relies on the effect of an earlier operation ($OpA(x)$ in the example) being delayed. 

For such a $h$, there are two possibilities.

\begin{enumerate}
\item[(i)] We can construct a program $P$ such that $P[C]$ has an observable behaviour not in $P[A]$ due to the delayed effect in $h$ (since a similar delayed effect is not possible in $P[A]$), i.e., $\exi P\dot \exi t:\semM{P[C]}\dot \res{t}{ir}=h \land \all t':\semM{P[A]}\dot \obsT{t'}\neq\obsT{t}$.
In this case, the antecedent of the theorem does not hold, and the theorem is trivially true.
\item[(ii)] We cannot construct such a program $P$ and hence $\all P\dot \all t:\semM{P[C]}\dot \res{t}{ir}=h \implies \exi t':\semM{P[A]}\dot \obsT{t'}=\obsT{t}$.
In this case, 
%
consider a program ${P_0}$ which allows the history $h$ to occur, and indicates (for any other history) when the operations occur in the same order as in $h$ by means of triangular races. For the example with $OpA$ and $OpB$, ${P_0}$ is of the form

\[z=1; || ...;u=OpA(x);w=z;... || ...; await(z=1);v=OpB(y);...\] 

where $z$ is initially 0 and the elided parts of the program do not change $z$, $w$ or $v$. If ${P_0}$ results in $w=0$, we can deduce that $OpA(x)$ has occurred before $OpB(y)$. Assume $OpB(y)$ in $C$ returns 1 when $OpA(x)$'s effect has taken place and 0 otherwise. 
For history $h$, the program $P_0[C]$ will additionally result in $v=0$ since the effect of $OpA(x)$ is delayed. Since we know there is an abstract trace $t'$ matching each concrete trace $t$ where $\res{t}{ir}=h$, it must be possible to have the result $w=v=0$ from ${P_0}[A]$. Therefore, there must be a history $h'\mem \sem{A}$ in which $OpA(x)$ occurs before $OpB(y)$ and the latter operation outputs 0 (even though the effect of $OpA(x)$ is not delayed). This history $h'$ is a linearization of $h$. Hence the consequent of the theorem holds for $h$. 

%
%
\end{enumerate}

For all other $h\mem \semM{C}$, a strongest program $\widehat{P}$ can be constructed. 
From Property~1, we know that $\widehat{P}$ allows $h$ to occur, i.e., there exists a trace of $\widehat{P}[C]$ corresponding to $h$, and that $\widehat{P}$ enforces the operations to occur in the order in $h$.
\begingroup
\renewcommand{\theequation}{C\thecompprops}
\begin{align}
&\exi t:\semM{\widehat{P}[C]}\dot \resIR{t}=h 
      \refstepcounter{compprops}\label{cp_1}\\
&\all c,d:Op\cross Val\dot (res(c),inv(d)) \mem \order{h}\implies (res(c),inv(d)) \mem\ <_{\widehat{P}_M}  
       \refstepcounter{compprops} \label{cp_3}
\end{align}
\endgroup

With Lemma~\ref{matchingTraces}  and the antecedent of the theorem we can deduce 
that if a concrete trace has a matching abstract trace there always exists another matching trace
which has the same invocations and responses.
\begingroup
\renewcommand{\theequation}{C\thecompprops}
\begin{align}
& \all t : \semM{\widehat{P}[C]} \dot \exi t'':\semM{\widehat{P}[A]} \dot \obsT{t}=\obsT{t''} 
                    \land \exi t^+:ext(t) \dot \resIR{comp(t^+)}\sim \resIR{t''}
       \refstepcounter{compprops} \label{cp_2}
\end{align}
\endgroup

\noindent
Let $t\mem \semM{\widehat{P}[C]}$ be any trace satisfying (\ref{cp_1}), i.e., 
\begin{c-equation}
\resIR{t}=h  \label{cp_2a}
\end{c-equation}
\!\!There exists an abstract trace $t''\in \semM{\widehat{P}[A]}$ 
and a trace $t^+\mem ext(t)$  such that the completion of $t^+$ and the abstract trace
share the same invocations and responses (from (\ref{cp_2})).
\begin{c-equation}
\resIR{comp(t^+)} \sim \resIR{t''}  \label{cp_13}
\end{c-equation}
\!\!With Lemma~\ref{subset}, it follows that
\begin{c-equation}
\supclass_{comp(t^+)}\ \subseteq\ \supclass_{t} \label{cp_6}
\end{c-equation}
Since $t'' \in \semM{\widehat{P}[A]}$, from definition of $\semM{\widehat{P}[A]}$ we have that $t'' \in \semM{\widehat{P}}$ and hence
\begin{c-equation}
events(t'') \subseteq events(\widehat{P})~ \land  ~
<_{\widehat{P}_M}\  \Subset \order{t''}  .\label{cp_11}
\end{c-equation}
and that there exists an $h'\mem \sem{A}$ such that
\begin{c-equation}
h'=\resIR{t''} \label{cp_16}
\end{c-equation}

\noindent
From (\ref{cp_3}) and (\ref{cp_2a}) we have 
\begin{c-equation}
\all c, d: Op \times Val \dot (res(c), inv(d)) \in \order{t} \implies (res(c), inv(d)) \in\ <_{\widehat{P}_M}  \label{cp_a}
\end{c-equation}
and with Lemma~\ref{subset} and the definition of $\supclass$
\begin{c-equation}
 \all c, d: Op \times Val, t^+ : ext(t) \dot (res(c), inv(d)) \in \order{comp(t^+)} \implies (res(c), inv(d)) \in \ <_{\widehat{P}_M}  \label{cp_b}
\end{c-equation}
Since $\Porder$ allows $t''$ (\ref{cp_11}) and with the definition of $\Subset$
\begin{c-equation}
\all c,d : Op\times Val \dot (res(c), inv(d)) \in \ <_{\widehat{P}_M} \ \land inv(d) \in events(t'') \implies (res(c),inv(d)) \in \order{t''} \label{cp_c}
\end{c-equation}

\noindent
From (\ref{cp_13}) we know that 
$ \all d: Op \times Val \dot inv(d) \in events(comp(t^+)) \implies inv(d) \in events(t'')$ and hence from (\ref{cp_b}) and (\ref{cp_c}) we have
\begin{c-equation}
\all c,d:Op\cross Val \dot
 (res(c),inv(d))\mem \order{comp(t^+)} \implies (res(c),inv(d))\mem \order{t''} \label{cp_d}
 \end{c-equation}
which is $\supclass_{comp(t^+)} \subseteq\, \supclass_{t''}$
and hence we also have
\begin{c-equation}
\supclass_{\resIR{comp(t^+)}} \subseteq\, \supclass_{\resIR{t''}} \label{cp_15}
\end{c-equation}
Since $t^+ \mem ext(t)$, and $h = \resIR{t}$, 
there exists a $h^+\mem ext(h)$ such that 
\begin{c-equation}
h^+ = \resIR{t^+} \label{cp_17}
\end{c-equation}
From (\ref{cp_17}) and (\ref{cp_15}), it follows that
\begin{c-equation}
\supclass_{comp(h^+)} \subseteq\, \supclass_{\resIR{t''}} \label{cp_100}
\end{c-equation}
and from (\ref{cp_16})
\begin{c-equation}
\supclass_{comp(h^+)} \subseteq\, \supclass_{h'} \label{cp_101}
\end{c-equation}
\!\!This is the second conjunct of the consequent. We now derive the first.\\

\noindent From (\ref{cp_16}) and  (\ref{cp_13}) we have
\begin{c-equation}
\resIR{comp(t^+)} \sim h' \label{cp_1021}
\end{c-equation}
and hence with (\ref{cp_17}) 
\begin{c-equation}
comp(h^+) \sim h' \label{cp_102}
\end{c-equation}
Thus conjoining (\ref{cp_102}) and (\ref{cp_101}) gives us
\begin{c-equation}
comp(h^+)\sim h' ~\land~ \supclass_{comp(h^+)} \subseteq \supclass_{h'} \label{cp_18}
\end{c-equation}
which is the consequent of the theorem.
\end{myproof}\vspace*{1ex}


\section{Example applications}
\label{sec:ex}

To demonstrate the application of standard linearizability for proving (or disproving) correctness of concurrent objects, 
we discuss two typical examples, a spinlock \cite{HeSh08} and a work-stealing deque \cite{LeWorkStealingPPoPP13} (the latter developed specifically for ARM). Our soundness result means that whenever an implementation is incorrect with respect to a specification on a given weak memory model, we can use linearizability to prove this. Our completeness result means that whenever an implementation is correct with respect to a specification on a given weak memory model, again we can use linearizability to prove it.

\subsection{Correctness on TSO}
\label{sec:tso}

Consider a spinlock object with operations {\sf acquire}, {\sf release} and {\sf tryAcquire} specified as follows.

\[ \parbox[t]{5cm}{\sf acquire\\ \hspace*{4mm}await(x=1);\\ \hspace*{4mm}x=0 
 } 
 \parbox[t]{4cm}{\sf release\\ \hspace*{4mm}x=1;
 } 
 \parbox[t]{4cm}{\sf tryAcquire\\ \hspace*{4mm}if (x=1) x=0; return 1 \\ \hspace*{4mm}else return 0} 
\]
%
\noindent A typical concurrent implementation \cite{HeSh08} which is correct on SC is 

\[
 \parbox[t]{6cm}{\sf acquire\\ \hspace*{4mm}while (true) $\{$\\  \hspace*{8mm}if\,(TAS(x,\,1,\,0)=1)\,return\\ 
\hspace*{8mm}while (x=0) $\{\}\\
 \hspace*{4mm}\}$
 } 
 \parbox[t]{3.4cm}{\sf release\\ \hspace*{4mm}x=1;
 } 
 \parbox[t]{3.8cm}{\sf tryAcquire\\ \hspace*{4mm}return\,TAS(x,\,1,\,0)}
\]
%
\noindent where {\sf TAS(x,a,b)} is the atomic hardware primitive
test-and-set which, when {\sf x} is {\sf a}, sets {\sf x} to {\sf b}
and returns~$1$, and otherwise returns $0$. The {\sf TAS} instruction has a built-in fence to ensure any change it makes to {\sf x} is immediately visible to all threads.

An earlier version of linearizability on TSO \cite{ifm} proved that this implementation is correct on TSO. However (as discussed in Section~\ref{sec:introduction}) it did not require that abstract operations take effect immediately. Under our definition of object refinement (for which we assume abstract operations take effect atomically), it is not correct. 

\begin{figure}
   \vspace{-8ex}
 \[\qquad \qquad\qquad\qquad\parbox[t]{2cm}{~~{\sf T1} \vspace*{2mm}\\ 
      {\sf z =  1;}} 
\parbox[h]{1cm}{\vspace*{2cm}{\resizebox{!}{0.5cm}{$\parallel$}}} 
\parbox[t]{2.5cm}{~~{\sf T2} \vspace*{2mm}\\ 
      {\sf sl.acquire();\\ sl.release();\\ y=z;}} 
     \parbox[h]{1cm}{\vspace*{2cm}{\resizebox{!}{0.5cm}{$\parallel$}}} 
      \parbox[t]{4cm}{~~{\sf T3}\vspace*{2mm}\\ {\sf await(z=1);\\ w = sl.tryAcquire();}}
 \]
 \caption{Program using spinlock}
  \label{fig:spinlock}
 \end{figure}

Consider the client program in Figure~\ref{fig:spinlock} which contains a triangular race. The program uses a spinlock object {\sf sl} for which we assume that initially {\sf x = 1} and {\sf z = 0}. Following the operational semantics of TSO in \cite{Sewell:2010:XRU:1785414.1785443}, one possible trace of this program is\footnote{The effects in TSO correspond to flushes for program steps and operations which write to global variables, and immediately follow the program step or operation response otherwise.} \\

\noindent$\hspace*{3mm}\lseq inv({\sf T2, sl.acquire), \bot}), res(({\sf T2, sl.acquire,\bot}), \obs(({\sf T2, sl.acquire), \bot}), inv(({\sf T2, sl.release),\bot}),$\\ 
$\hspace*{3mm}res({\sf T2, sl.release,\bot}), step({\sf T2, y=0}), step({\sf T1, z=1)}, \obs({\sf T1, z=1}), step({\sf T3, await(z=1)}),$ \\
$\hspace*{3mm}\obs({\sf T3, await(z=1)}), inv(({\sf T3, sl.tryAcquire), \bot}), res(({\sf T3, sl.tryAcquire),0}), \obs(({\sf T3, sl.tryAcquire),0}),$\\
$\hspace*{3mm}step({\sf T3, w=0}), \obs({\sf T3, w=0}), \obs(({\sf T2, sl.release),\bot}), \obs({\sf T2, y=0})\rseq$\\

\noindent This trace corresponds to thread {\sf T2} acquiring and releasing the lock and reading the initial value of {\sf z}, but not flushing the value written to {\sf x} by the {\sf release} operation until after the other two threads have run to completion. The observable behaviour of the trace is \\

$\lseq  \obs({\sf T1, z=1}), \obs({\sf T3, w=0)}, \obs({\sf T2,y=0})\rseq$\\
~\\
This is not an observable trace of the program running with an abstract object whose effects are not delayed: if {\sf y=0} then the step {\sf y=z} and hence {\sf sl.release} on {\sf T2} must have occurred before {\sf z=1} on {\sf T1}, and hence before {\sf sl.tryAcquire} on {\sf T3}. Hence, we do not have object refinement. We can prove this with the standard definition of linearizability due to it being sound with respect to object refinement. 

The spinlock implementation without the {\sf tryAcquire} operation is, however, known to be correct on TSO
\cite{Sewell:2010:XRU:1785414.1785443}. Again we can show this using our definition of object refinement. 

The traces of the implementation derived from the operational semantics of TSO show that if an {\sf acquire} has responded (and hence has been observed due to the fence in the {\sf TAS}) then another {\sf acquire} cannot respond until after a {\sf release} on the same core has responded or a {\sf release} on another core has been observed. 
This coincides with what can be observed from the abstract specification, i.e., an {\sf acquire} is always followed by a {\sf release} (before another {\sf acquire} can occur). Hence, object refinement holds. Again we can prove this with the standard definition of linearizability due to it being complete with respect to object refinement.  

\subsection{Correctness on ARM and POWER}
\label{sec:arm}


It is easy to show that the spinlock implementation of Section~\ref{sec:tso},
even without the {\sf tryAcquire} operation, 
is not correct on ARM and POWER using our definition of object refinement.
For example, consider the client program in Figure~\ref{fig:client-arm} for which we assume 
that initially {\sf x = 1} and {\sf y = 0}.
Following the operational semantics of ARM and POWER given in \cite{ColvinFM2018},
one possible trace of this program is\footnote{In the operational semantics of \cite{ColvinFM2018} the placement of effects can be derived from the model of the ``storage subsystem'' which keeps track of which updates to global variables have been seen by which threads.}\\

\begin{figure}
  \vspace{-8ex}
\[
 \qquad \qquad \qquad \quad \qquad\qquad\parbox[t]{2.6cm}{~~{\sf T1} \vspace*{2ex}\\ 
  {\sf sl.acquire();\\ y=y + 1; \\ sl.release()}} 
    \parbox[h]{1cm}{\vspace*{2.6cm}{\resizebox{!}{0.5cm}{$\parallel$}}} 
    \parbox[t]{3cm}{~~{\sf T2}\vspace*{2ex}\\ {\sf sl.acquire();\\ y=y + 1;\\sl.release()}}
\]
\caption{Another program using spinlock}
 \label{fig:client-arm}
\end{figure}

\noindent
$ \lseq 
  inv({\sf (T1, sl.acquire), \bot}), res({\sf (T1, sl.acquire), \bot}), \obs({\sf (T1, sl.acquire), \bot}),
   step({\sf T1, y = 1}), $\\
$   inv({\sf (T1, sl.release), \bot}), res({\sf (T1, sl.release), \bot}), \obs({\sf (T1, sl.release), \bot}), 
   inv({\sf (T2, sl.acquire), \bot}),$\\
   $res({\sf (T2, sl.acquire), \bot}),
   \obs({\sf (T2, sl.acquire), \bot}),
   step({\sf T2, y = 1}), inv({\sf (T2, sl.release), \bot}), res({\sf (T2, sl.release), \bot}), $\\
   $\obs({\sf (T2, sl.release), \bot}), \obs({\sf T1, y=1}), \obs({\sf T2, y = 1})\rseq
$\\
~\\
This trace corresponds to the response of {\sf T1}'s {\sf release} operation being observed before its update to~{\sf y}. This allows {\sf T2}'s {\sf acquire} to occur followed by its update of {\sf y} before {\sf T1}'s new value of {\sf y} is observable by {\sf T2}. Hence, both threads update {\sf y} to 1. 

Since the observable behaviour $\lseq \obs({\sf T1, y=1}), \obs({\sf
  T2, y=1})\rseq$ of the above trace is not possible using the specification,
the implementation is not correct on ARM or POWER: object refinement does not hold. We can prove this with the standard definition of linearizability due to it being sound with respect to object refinement.  

\subsubsection{Chase-Lev deque}

As a more substantial example, consider the below code 
for a version of the Chase-Lev work-stealing deque (double-ended queue) \cite{ChaseLev05} developed specifically for ARM \cite{LeWorkStealingPPoPP13}. The code shown corresponds to a refactoring used in \cite{ColvinFM2018} which, for example, eliminates returns from within a branch, and for simplicity assumes the elements of the deque are integers.

\[ \parbox[t]{4.5cm}{\sf put(v)\\
\hspace*{4mm}int t;\\
\hspace*{4mm}t=tail;\\
\hspace*{4mm}tasks[t mod L]=v;\\
\hspace*{4mm}fence;\\
\hspace*{4mm}tail=t+1
 } 
 \parbox[t]{6.5cm}{\sf take\\ 
 \hspace*{4mm}int h,t,task;\\
 \hspace*{4mm}t=tail-1;\\
 \hspace*{4mm}tail=t;\\
 \hspace*{4mm}fence;\\
 \hspace*{4mm}h=head;\\
 \hspace*{4mm}if (h $<$= t)  \\
 \hspace*{8mm}task=tasks[t mod L];\\
 \hspace*{8mm}if (h=t)\\
 \hspace*{12mm}if !CAS(head, h, h + 1) then\\
 \hspace*{16mm}task=empty;\\
 \hspace*{12mm}tail=tail+1;\\
 \hspace*{4mm}else\\
 \hspace*{8mm}task=empty;\\
 \hspace*{8mm}tail=tail+1;\\
\hspace*{4mm}return task \\
} 
\parbox[t]{5cm}{\sf steal\\ 
\hspace*{4mm}int h,t,task; \\
\hspace*{4mm}h=head; \\
\hspace*{4mm}fence; \\
\hspace*{4mm}t=tail; \\
\hspace*{4mm}cfence; \\
\hspace*{4mm}if (h $<$ t) \\
\hspace*{8mm}task=tasks[h mod L]; \\
\hspace*{8mm}cfence; \\ 
\hspace*{8mm}if !CAS(head, h, h+1)  \\ 
\hspace*{12mm}task=fail; \\ 
\hspace*{4mm}else \\ 
\hspace*{8mm}task=empty;\\
\hspace*{4mm}return task \\
}
\]

The deque is implemented as a circular array of size {\sf L} with a {\sf head} and {\sf tail} pointer. Elements may be \emph{put} on or \emph{taken} from the tail by a worker thread, and additionally, other (thief) processes may \emph{steal} an element from the head of the deque (in order to balance system workload).
Since the {\sf put} and {\sf take} operations
are executed by a single thread, there is no interference between these two operations. 

The {\sf put} operation straightforwardly adds an element to the end of the deque, incrementing the {\sf tail} pointer. It includes a full fence so that the increment of the tail pointer does not take effect before the element is placed in the array.  

The interesting behaviour is in the way that the {\sf take} and {\sf steal} operations interact when called concurrently. To take the task at position {\sf t=tail-1}, the worker process decrements {\sf tail} to equal {\sf t}, thereby publishing its intent to take that task. This publication, ensured by the following fence, means subsequent thief processes will not try to steal the task at position {\sf t}. It then reads {\sf head} into {\sf h} and if {\sf h $<$ t} knows that there is more than one task in the deque and it is safe to take the task at position {\sf t}, i.e., no thief process can concurrently steal it.

If {\tt t < h} the worker knows the deque is empty and sets {\tt tail} back to its original value. The final possibility is that {\tt h=t}. In this case, there is one task on the deque and conflict with a thief may arise. To deal with this conflict, both the {\tt take} and {\tt steal} operations employ an atomic CAS (compare-and-swap) operation. An operation {\tt CAS(x,y,z)} checks whether {\tt x} equals {\tt y} and, if so, updates {\tt x} to {\tt z} and returns true, otherwise it returns false leaving {\tt x} unchanged. The CAS is atomic, and the update is immediately written to memory since the CAS operation also implements a fence. 

 If {\tt h=t}, rather than decrementing {\tt tail} to take the task, the worker increments {\tt head}. Therefore, if the worker finds {\tt h=t}, it restores {\tt tail} to its original value. The {\tt steal} operation works similarly. The operation reads the deque's head and tail into {\tt h} and {\tt t}, and if the deque is not empty tries to increment {\tt head} from {\tt h} to {\tt h+1} using a CAS. If it succeeds, the value of {\tt head} has not been changed since read into the local variable {\tt h} and hence the thief has stolen the task. 

Additionally, the {\sf steal} operation contains two control fence barriers ({\tt ctrl\_isync} in ARM), denoted {\sf cfence} in code above. A control fence ensures all branch instructions occurring before it take effect before any {\em loads\/}, i.e., reads of global variables, occurring after it. It can therefore be used to avoid {\em speculative execution\/} of loads occurring in a branch. As shown in \cite{ColvinFM2018},  the first control fence is redundant, and the second is incorrectly placed.  Eliminating the first {\sf cfence} and swapping the order of the second {\sf cfence} with the preceding assignment gives the intended behaviour.

To show that we could also detect these problems using standard linearizability, we define the specification of the expected deque behaviour in terms of a sequence $q$ and sequence operations $\cat$ (sequence concatentation), {\sf head}, {\sf tail}, {\sf last} and {\sf front}. We also use $\sqcap$ to denote nondeterministic choice, which is required to model that {\sf steal} may fail.

\[ \parbox[t]{4.5cm}{\sf put(v)\\
\hspace*{4mm}q=q$\cat \lseq$ v$\rseq$
 } 
 \parbox[t]{6.5cm}{\sf take\\ 
 \hspace*{4mm}if q=$\emptyseq$ \\
\hspace*{8mm}return empty\\
 \hspace*{4mm}else\\
\hspace*{8mm}q=front(q);\\
\hspace*{8mm}return last(q)
} 
\parbox[t]{5cm}{\sf steal\\ 
 \hspace*{4mm}if q=$\emptyseq$ \\
\hspace*{8mm}return empty\\
 \hspace*{4mm}else\\
\hspace*{8mm}q=tail(q); return head(q)\\
\hspace*{8mm}$\sqcap$\\
\hspace*{8mm}return fail
}
\]

\begin{figure}
  \vspace{-8ex}
\[
 \qquad \qquad \qquad\qquad \qquad \quad \qquad\qquad\parbox[t]{2.6cm}{~~{\sf T1} \vspace*{2ex}\\ 
  {\sf d.put(1);}} 
    \parbox[h]{1cm}{\vspace*{1.3cm}{\resizebox{!}{0.5cm}{$\parallel$}}} 
    \parbox[t]{3cm}{~~{\sf T2}\vspace*{2ex}\\ {\sf y:=d.steal()}}
\]
\caption{Program using the Chase-Lev deque}
 \label{fig:cl1}
\end{figure}

Consider the simple program in Figure~\ref{fig:cl1} in which the deque $d$ is initially empty and the circular buffer implementing it initially holds only zeroes. An example trace of the program is

\[\lseq 
  inv({\sf (T2, d.steal), \bot}), inv({\sf (T1, d.put), 1}), res({\sf (T1, d.put), \bot}), \obs({\sf (T1, d.put), \bot}),\obs({\sf (T2, d.steal), 0}),\\
 res({\sf (T2, d.steal), 0}), step({\sf T2,y=0}), \obs({\sf T2,y=0})\rseq
\]
This trace can occur since the load, {\sf t=tail}, in {\sf steal} does not need to take effect before the first control fence, and the load after the {\sf if} instruction can take effect before the {\sf if} statement, so that the code between the full fence and the second control fence is effectively

\[{\sf cfence; t=tail; task=tasks[h ~mod~ L]; if (h < t)}\ .\]
Since the two loads are also not required to take effect in order on ARM, this can be effectively executed as 

\[{\sf cfence; task=tasks[h ~mod~ L]; t=tail; if (h < t)}\ .\]

Now assume that the {\sf steal} operation is invoked and reaches the load, {\sf task=tasks[h mod L]}. Since nothing has been put into the deque, {\sf task} will be set to the initial value at {\sf h} which is 0. If the {\sf put} operation is then invoked and takes effect, when the {\sf steal} continues it will load the new value of {\sf tail} into {\sf t}. Since this new value is greater than {\sf h}, and no other operation will change {\sf h} in the program of Figure~\ref{fig:cl1}, the {\sf steal} operation will complete returning 0.

The observable behaviour of this trace 

\[\lseq eff({\sf T2, y=0})\rseq\]
is not a possible observable trace of the program running with an abstract object: the only observable traces are $\lseq eff({\sf T2, y=empty})\rseq$ and $\lseq eff({\sf T2, y=1})\rseq$. Hence, the failure of the implementation to meet the specification could have been detected using standard linearizability.

As suggested in \cite{ColvinFM2018}, the implementation can be fixed by moving the second control fence before the load into {\sf task}. 
This ensures the load into {\sf task} takes effect after the {\sf if} statement, and hence also after the load into {\sf t} (the latter cannot take effect after the {\sf if} statement according to the semantics in \cite{ColvinFM2018}). Hence, {\sf task} in {\sf steal} can only be set to {\sf empty} or 1, and the only observable behaviours of the program are those satisfying the specification.  

\begin{figure}
   \vspace{-8ex}
 \[\qquad \qquad\qquad\qquad\qquad\parbox[t]{2cm}{~~{\sf T1} \vspace*{2mm}\\ 
      {\sf z =  1;}} 
\parbox[h]{1cm}{\vspace*{1.5cm}{\resizebox{!}{0.5cm}{$\parallel$}}} 
\parbox[t]{2.5cm}{~~{\sf T2} \vspace*{2mm}\\ 
      {\sf d.put(1);\\ y=z;}} 
     \parbox[h]{1cm}{\vspace*{1.5cm}{\resizebox{!}{0.5cm}{$\parallel$}}} 
      \parbox[t]{4cm}{~~{\sf T3}\vspace*{2mm}\\ {\sf await(z=1);\\ w = d.steal();}}
 \]
 \caption{Another program using the Chase-Lev deque}
  \label{fig:cl2}
 \end{figure}

Consider, however, the program containing a triangular race in Figure~\ref{fig:cl2}. On ARM the {\sf await} instruction includes a fence to prevent subsequent loads taking effect earlier than it. A possible behaviour of this program when executed with the fixed implementation is\\

\noindent$\hspace*{3mm}\lseq inv({\sf T2, d.put), \bot}), res({\sf T2, d.put,\bot}), step({\sf T2, y=0}), step({\sf T1, z=1)}, \obs({\sf T1, z=1}), step({\sf T3, await(z=1)}),$ \\
$\hspace*{3mm}\obs({\sf await(z=1)}), inv(({\sf T3, d.steal), \bot}), res(({\sf T3, d.steal),empty}), \obs(({\sf T3, d.steal),empty}),  $\\
$\hspace*{3mm}step({\sf T3, w=empty}), \obs({\sf T3, w=empty}),\obs(({\sf T2, d.put),\bot}), \obs({\sf T2, y=0})\rseq$\\
~\\
This occurs when the effect of the {\sf put} operation is delayed until after the {\sf steal} operation has occurred. Specifically, {\sf tail} is not updated until after the {\sf steal} operation. The observable behaviour is 

\[ \lseq  \obs({\sf T1, z=1}), \obs({\sf T3, w=empty)}, \obs({\sf T2,y=0})\rseq\]
which is not a possible behaviour according to the specification. Hence, this would be detected when attempting to prove standard linearizability. It could be fixed by adding a fence at the end of the {\sf put} operation. Without this additional fence, the program of Figure~\ref{fig:cl2} can produce the unexpected observable behaviour above. If this behaviour is deemed acceptable (as it is harmless if a thief process returns {\sf empty} when there is in fact an element in the deque) then to prove correctness would require a modification of the specification to allow {\sf steal} to nondeterministically return {\sf empty} in such circumstances.  \\

The need for a fence at the end of the {\sf put} operation highlights a consequence of our notion of correctness. As shown in Figure~\ref{fig:cl2}, any operation whose effect can influence the outcome of another operation needs to be fenced to be correct on a weak memory model. Note however that this does not mean all operations require fences. One example of an operation where a fence is not required is the {\sf release} operation of spinlock on TSO. Also, both the {\sf take} and {\sf steal} operations of the Chase-Lev deque on ARM do not need to be fenced when they fail to return a task. In particular, it is not a problem when {\sf take} does not immediately restore  {\sf tail} to its original value before the next {\sf steal}. Restoring {\sf tail}'s original value is only required in {\sf take} when the deque is empty, and if another {\sf put} occurs then {\sf tail} will be correctly updated by the {\sf fence} in {\sf put} (since {\sf put} and {\sf take} are both only called by the same worker thread).

\section{Related work}
\label{sec:relwork}

Traditionally, trace refinement provides the notion of correctness
for programs: an implementation is correct with respect to its
specification if and only if each observable behaviour of the
implementation can also be observed from the specification \cite{bac94a,back,abadi}. 
In recent work \cite{Smith2017} we showed that trace refinement and linearizability do not coincide; linearizability is, in general, weaker than trace refinement. This is because in the context of concurrent objects that are called by a client program, 
an object implementation is deemed correct if and only if the client program cannot differentiate between 
the object implementation and its abstract specification. This notion of correctness differs from trace refinement.

To capture what is observed by client programs under \emph{sequentially consistent} (SC) architectures (i.e., those without a weak memory model), 
the notions of \emph{observational refinement} \cite{Filipovic-LinvsRef2010} and 
\emph{contextual refinement} \cite{Dongol2016} have been
introduced. However, both of these definitions do not provide a notion
of correctness under weak memory models, as they do not take into account
that an event occurring on one thread might be observable later on another. Instead
events become observable immediately after their occurrence. Our definition of object refinement provides a correctness notion for concurrent objects which is applicable to weak memory models.

Bouajjani et al.\ \cite{BouajjaniMeyer_2011,BouajjaniMeyer_2013} introduce a notion of \emph{robustness} which requires each trace of a concurrent implementation running on a weak memory model (specifically, TSO) to be equivalent to a trace of the implementation running on a sequentially consistent (SC) architecture. This is a stronger requirement than object refinement which requires that each trace of a concurrent implementation has the same \emph{observable behaviour} as a trace of a specification. This allows the implementation to exhibit non-SC behaviour provided that it cannot be observed by any client program. The {\sf release} operation of spinlock on TSO, and the {\sf take} and {\sf steal} operations of the Chase-Lev deque on ARM provide examples of where non-SC behaviour is allowed. Furthermore, robustness checks a correspondence between a given implementation on two different memory models (a weak memory model and SC). Object refinement, on the other hand, checks a correspondence between an implementation and a specification. If the specification is sufficiently weak then the implementation may exploit non-SC behaviour to implement it. 
\medskip

Two recent publication are related to our result: 
Dongol et al.\ \cite{DongolVMCAI2018}  investigate the effects of hardware memory models like
TSO, ARM and POWER, and 
Doherty et al.\ \cite{DohertyIFM2018}  investigate linearizability for concurrent objects under C11.

Dongol et al.\ \cite{DongolVMCAI2018} define {\em real-time
  hb-linearisability\/} as the correctness condition for concurrent objects running on multicore processors.
This differs from our result due to the way the semantics
of object implementations and specifications are characterised. Similar to our work, traces follow a real-time ordering of events which (on one thread) corresponds to the program order.
Additionally, however, the events in each trace (including the output values) have to satisfy the happens-before ordering \emph{hb}
(as defined in \cite{HerdingCats}) postulated for the weak memory model. For the abstract
specification an additional \emph{specification order} is introduced that ensures the intended
sequences of events (and output values). 
\emph{Real-time hb-linearisability} extends standard linearizability with an additional check
that the specification order is maintained between matching abstract and concrete histories.

In our work, both the happens-before order and the specification order are incorporated into
the semantics of concrete and abstract objects. Instead of introducing
an additional check in the linearizability definition (and hence strengthening it), 
we abstract from the ordering constraints of $M$ and assume that $\sem{A}$ 
contains only those histories that satisfy the specification order (and produce the intended output values)
and similarly $\semM{C}$ contains only
histories in which sequences of events and their output values satisfy the constraints imposed 
by the object's code and the memory model. 

 The benefits of this viewpoint are threefold:
Firstly we can observe that standard linearizability is sound and complete with respect to object 
refinement (no strengthening is required), 
and standard linearizability is known to be compositional \cite{HeWi90}. 
Secondly, our semantic model is generic and hence independent of any particular weak memory model semantics
(whereas \cite{DongolVMCAI2018} is linked to the semantic model of \cite{HerdingCats}).
Thirdly, the assumptions we make on program/object semantics under weak memory models
are formalised as simple axioms (see Sections~\ref{sec:progsemantics} and \ref{sec:objectsemantics}). 
These axioms are generic in that they are based on principles of object encapsulation
and abstract specification
-- no reasoning about weak memory behaviour is required at this level.

Dongol et al.\ also define {\em causal hb-linearisability\/} for processes running on distributed systems (and potentially future multicore architectures). This
definition, motivated by requiring programs to have less
synchronisation overhead, removes the need to preserve the real-time
order on events between the abstract and concrete specifications,
preserving instead the happens-before order. 
As outlined in \cite{DongolVMCAI2018} this more relaxed
definition allows us to prove objects such as that in our Figure~\ref{fig:unsound} (repeated below) linearizable. 
 \vspace{-2mm}
 \[
   \qquad \qquad\qquad\qquad\qquad~~~~
    \parbox[t]{2.1cm}{{\sf o.WriteX;\\ z=o.ReadY}} 
    \parbox[h]{1cm}{\vspace*{1.5ex}{\resizebox{!}{0.5cm}{$\parallel$}}} 
    \parbox[t]{3cm}{{\sf o.WriteY;\\ w=o.ReadX}}
 \]

As discussed in Section~\ref{sec:introduction} this program can produce the result ({\sf z=w=0}) on TSO when the {\sf WriteX} and {\sf WriteY} operations write value $1$ to variable $x$ and value $1$ to variable $y$, respectively. This
does not match any behaviour of the abstract object 
whose operations are viewed as taking effect immediately.
Hence causal hb-linearisability is not sound with respect to our definition of object refinement. 

Doherty et al.\ \cite{DohertyIFM2018} define {\em causal linearizability\/} as their correctness condition
which is closely related to causal hb-linearisability \cite{DongolVMCAI2018} but improves on its
compositionality result. Whilst causal hb-linearisability is only compositional under certain constraints 
either on the abstract specification or on the client program, causal linearizabiltiy is proved to be compositional for any specification and under any client. 
As with causal hb-linearisability, however, it does not target hardware weak memory models, but the C11 memory model. Consequently, it also is not sound with respect to our definition of object refinement.
\vspace*{4ex}

\noindent{\bf Acknowledgement} The authors would like to thank Lindsay Groves for his valuable feedback and advice. This work was supported by Australian Research Council Discovery Grant DP160102457.

\bibliographystyle{alpha}
\bibliography{references}

\newcommand{\etalchar}[1]{$^{#1}$}
\begin{thebibliography}{MHMS{\etalchar{+}}12}

\bibitem[AFI{\etalchar{+}}08]{Alglave:2009:SPA:1481839.1481842}
J.~Alglave, A.~Fox, S.~Ishtiaq, M.~O. Myreen, S.~Sarkar, P.~Sewell, and F.Z.
  Nardelli.
\newblock {The Semantics of Power and ARM Multiprocessor Machine Code}.
\newblock In L.~Petersen and M.M.T. Chakravarty, editors, {\em DAMP '09}, pages
  13--24. ACM, 2008.

\bibitem[AL91]{abadi}
M.~Abadi and L.~Lamport.
\newblock The existence of refinement mappings.
\newblock {\em Theoretical Computer Science}, 82(2):253--284, 1991.

\bibitem[AMT14]{HerdingCats}
J.~Alglave, L.~Maranget, and M.~Tautschnig.
\newblock Herding cats: Modelling, simulation, testing, and data mining for
  weak memory.
\newblock {\em ACM Trans. Program. Lang. Syst.}, 36(2):7:1--7:74, 2014.

\bibitem[Bac90]{back}
R.-J.R. Back.
\newblock Refinement calculus, part {II}: Parallel and reactive programs.
\newblock In {\em Stepwise Refinement of Distributed Systems Models,
  Formalisms, Correctness}, pages 67--93. Springer, 1990.

\bibitem[BDM13]{BouajjaniMeyer_2013}
A.~Bouajjani, E.~Derevenetc, and R.~Meyer.
\newblock Checking and enforcing robustness against {TSO}.
\newblock In M.~Felleisen and P.~Gardner, editors, {\em Programming Languages
  and Systems (ESOP 2013)}, pages 533--553. Springer, 2013.

\bibitem[BGMY12]{bur12}
S.~Burckhardt, A.~Gotsman, M.~Musuvathi, and H.~Yang.
\newblock Concurrent library correctness on the {TSO} memory model.
\newblock In H.~Seidl, editor, {\em ESOP 2012}, volume 7211 of {\em LNCS},
  pages 87--107. Springer, 2012.

\bibitem[BMM11]{BouajjaniMeyer_2011}
A.~Bouajjani, R.~Meyer, and E.~M{\"o}hlmann.
\newblock Deciding robustness against total store ordering.
\newblock In Luca Aceto, Monika Henzinger, and Ji{\v{r}}{\'i} Sgall, editors,
  {\em Automata, Languages and Programming}, pages 428--440. Springer, 2011.

\bibitem[BOS{\etalchar{+}}11]{BattyOSSW11}
M.~Batty, S.~Owens, S.~Sarkar, P.~Sewell, and T.~Weber.
\newblock Mathematizing {C++} concurrency.
\newblock In {\em POPL}, pages 55--66. {ACM}, 2011.

\bibitem[BvW94]{bac94a}
R.-J.R. Back and J.~von Wright.
\newblock Trace refinement of action systems.
\newblock In {\em CONCUR '94}, volume 836 of {\em LNCS}, pages 367--384.
  Springer, 1994.

\bibitem[CL05]{ChaseLev05}
D.~Chase and Y.~Lev.
\newblock Dynamic circular work-stealing deque.
\newblock In {\em SPAA'05: Proceedings of the 17th annual ACM symposium on
  Parallelism in algorithms and architectures}, pages 21--28, New York, NY,
  USA, 2005. ACM Press.

\bibitem[CS18]{ColvinFM2018}
R.J. Colvin and G.~Smith.
\newblock A wide-spectrum language for verification of programs on weak memory
  models.
\newblock In K.~Havelund, J.~Peleska, B.~Roscoe, and E.~de~Vink, editors, {\em
  {FM 2018}}, volume 10951 of {\em LNCS}, pages 240--257. Springer, 2018.

\bibitem[DD16]{DohertyDerrick2016}
S.~Doherty and J.~Derrick.
\newblock Linearizability and causality.
\newblock In {\em SEFM 2016}, volume 9763 of {\em LNCS}, pages 45--60.
  Springer, 2016.

\bibitem[DDGS15]{Dongol2015a}
B.~Dongol, J.~Derrick, L.~Groves, and G.~Smith.
\newblock Defining correctness conditions for concurrent objects in multicore
  architectures.
\newblock In {\em ECOOP '15}, LIPIcs, pages 470--494. Schloss Dagstuhl --
  Leibnis-Zentrum f{\"u}r Informatik, 2015.

\bibitem[DDWD18]{DohertyIFM2018}
S.~Doherty, B.~Dongol, H.~Wehrheim, and J.~Derrick.
\newblock Making linearizability compositional for partially ordered
  executions.
\newblock In C.~A. Furia and K.~Winter, editors, {\em Integrated Formal Methods
  (iFM 2018)}, volume 11023 of {\em Lecture Notes in Computer Science}, pages
  110--129. Springer, Cham, 2018.

\bibitem[DG16]{Dongol2016}
B.~Dongol and L.~Groves.
\newblock Contextual trace refinement for concurrent objects: Safety and
  progress.
\newblock In K.~Ogata, M.~Lawford, and S.~Liu, editors, {\em ICFEM 2016}, pages
  261--278. Springer, 2016.

\bibitem[DJRA18]{DongolVMCAI2018}
B.~Dongol, R.~Jagadeesan, J.~Riely, and A.~Armstrong.
\newblock On abstraction and compositionality for weak-memory linearisability.
\newblock In I.~Dillig and J.~Palsberg, editors, {\em {VMCAI}'18}, volume 10747
  of {\em LNCS}, pages 183--204. Springer, 2018.

\bibitem[DS15]{DBLP:conf/fm/DerrickS15}
J.~Derrick and G.~Smith.
\newblock A framework for correctness criteria on weak memory models.
\newblock In N.~Bj{\o}rner and F.S. de~Boer, editors, {\em {FM} 2015}, volume
  9109 of {\em LNCS}, pages 178--194. Springer, 2015.

\bibitem[DSD14]{ifm}
J.~Derrick, G.~Smith, and B.~Dongol.
\newblock Verifying linearizability on {TSO} architectures.
\newblock In E.~Albert and E.~Sekerinski, editors, {\em iFM 2014}, volume 8739
  of {\em LNCS}, pages 341--356. Springer, 2014.

\bibitem[DSGD14]{der14}
J.~Derrick, G.~Smith, L.~Groves, and B.~Dongol.
\newblock Using coarse-grained abstractions to verify linearizability on {TSO}
  architectures.
\newblock In E.~Yahav, editor, {\em HVC 2014}, pages 1--16. Springer, 2014.

\bibitem[DSGD17]{Derrick2017}
J.~Derrick, G.~Smith, L.~Groves, and B.~Dongol.
\newblock A proof method for linearizability on {TSO} architectures.
\newblock In M.~Hinchey, J.P. Bowen, and E.-R. Olderog, editors, {\em Provably
  Correct Systems}, pages 61--91. Springer, 2017.

\bibitem[DSW11]{DSW10mvpolin}
J.~Derrick, G.~Schellhorn, and H.~Wehrheim.
\newblock Mechanically verified proof obligations for linearizability.
\newblock {\em ACM Trans. Program. Lang. Syst.}, 33(1):4:1--4:43, 2011.

\bibitem[FGP{\etalchar{+}}16]{armv8}
S.~Flur, K.E. Gray, C.~Pulte, S.~Sarkar, A.~Sezgin, L.~Maranget, W.~Deacon, and
  P.~Sewell.
\newblock Modelling the {ARMv8} architecture, operationally: Concurrency and
  {ISA}.
\newblock In R.~Bodik and R.~Majumdar, editors, {\em POPL 2016}, pages
  608--621. ACM, 2016.

\bibitem[FORY10]{Filipovic-LinvsRef2010}
I.~Filipovi\'c, P.~W. O'Hearn, N.~Rinetzky, and H.~Yang.
\newblock Abstraction for concurrent objects.
\newblock {\em Theoretical Computer Science}, 411(51-52):4379 -- 4398, 2010.

\bibitem[GMY12]{got12}
A.~Gotsman, M.~Musuvathi, and H.~Yang.
\newblock Show no weakness: Sequentially consistent specifications of {TSO}
  libraries.
\newblock In M.~Aguilera, editor, {\em DISC 2012}, volume 7611 of {\em LNCS},
  pages 31--45. Springer, 2012.

\bibitem[GY11]{GotsmanYang2011}
A.~Gotsman and H.~Yang.
\newblock Liveness-preserving atomicity abstraction.
\newblock In {\em ICALP 2011}, volume 6756 of {\em LNCS}, pages 453--465.
  Springer, 2011.

\bibitem[HS08]{HeSh08}
M.~Herlihy and N.~Shavit.
\newblock {\em \hspace*{-0.5mm}The Art of Multiprocessor Programming}.
\newblock \hspace*{-0.5mm}Morgan Kaufmann, \hspace*{-0.6mm}2008.

\bibitem[HW90]{HeWi90}
M.~Herlihy and J.~M. Wing.
\newblock Linearizability: A correctness condition for concurrent objects.
\newblock {\em ACM Trans. Program. Lang. Syst.}, 12(3):463--492, 1990.

\bibitem[LPCZN13]{LeWorkStealingPPoPP13}
N.M. L\^{e}, A.~Pop, A.~Cohen, and F.~Zappa~Nardelli.
\newblock Correct and efficient work-stealing for weak memory models.
\newblock In {\em Proceedings of the 18th ACM SIGPLAN Symposium on Principles
  and Practice of Parallel Programming}, PPoPP '13, pages 69--80, New York, NY,
  USA, 2013. ACM.

\bibitem[MHMS{\etalchar{+}}12]{AxiomaticPOWER}
S.~Mador-Haim, L.~Maranget, S.~Sarkar, K.~Memarian, J.~Alglave, S.~Owens,
  R.~Alur, M.M.K. Martin, P.~Sewell, and D.~Williams.
\newblock An axiomatic memory model for {POWER} multiprocessors.
\newblock In {\em CAV'12}, pages 495--512, Berlin, Heidelberg, 2012.
  Springer-Verlag.

\bibitem[MS04]{Moi07}
M.~Moir and N.~Shavit.
\newblock Concurrent data structures.
\newblock {\em Handbook of Data Structures and Applications}, pages
  47:1--47:30, 2004.

\bibitem[NMS16]{NienhuisMS16}
K.~Nienhuis, K.~Memarian, and P.~Sewell.
\newblock An operational semantics for {C/C++11} concurrency.
\newblock In {\em OOPSLA}, pages 111--128. {ACM}, 2016.

\bibitem[Owe10]{DBLP:conf/ecoop/Owens10}
S.~Owens.
\newblock Reasoning about the implementation of concurrency abstractions on
  {x86-TSO}.
\newblock In T.~D'Hondt, editor, {\em ECOOP 2010}, volume 6183 of {\em LNCS},
  pages 478--503. Springer, 2010.

\bibitem[SSA{\etalchar{+}}11]{UnderstandingPOWER}
S.~Sarkar, P.~Sewell, J.~Alglave, L.~Maranget, and D.~Williams.
\newblock Understanding {POWER} multiprocessors.
\newblock {\em SIGPLAN Not.}, 46(6):175--186, June 2011.

\bibitem[SSO{\etalchar{+}}10]{Sewell:2010:XRU:1785414.1785443}
P.~Sewell, S.~Sarkar, S.~Owens, F.Z. Nardelli, and M.O. Myreen.
\newblock {x86-TSO: a rigorous and usable programmer's model for x86
  multiprocessors}.
\newblock {\em Commun. ACM}, 53(7):89--97, 2010.

\bibitem[SW17]{Smith2017}
G.~Smith and K.~Winter.
\newblock Relating trace refinement and linearizability.
\newblock {\em Formal Aspects of Computing}, 29(6):935--950, 2017.

\bibitem[SWC18]{refine18}
G.~Smith, K.~Winter, and R.J. Colvin.
\newblock Correctness of concurrent objects under weak memory models.
\newblock In J.~Derrick, B.~Dongol, and S.~Reeves, editors, {\em Refine 2018},
  volume 282 of {\em EPTCS}, pages 53--67. Open Publishing Association, 2018.

\bibitem[TMW13]{DBLP:conf/hvc/TravkinMW13}
O.~Travkin, A.~M{\"u}tze, and H.~Wehrheim.
\newblock {SPIN} as a linearizability checker under weak memory models.
\newblock In V.~Bertacco and A.~Legay, editors, {\em HVC2013}, volume 8244 of
  {\em LNCS}, pages 311--326. Springer, 2013.

\bibitem[TW14]{hvc14}
O.~Travkin and H.~Wehrheim.
\newblock Handling {TSO} in mechanized linearizability proofs.
\newblock In E.~Yahav, editor, {\em HVC2014}, volume 8855 of {\em LNCS}, pages
  132--147. Springer, 2014.

\end{thebibliography}

\newpage
\appendix

\newenvironment{lemmaproof}[1]{\par\noindent\textbf{Lemma \ref{#1}.}}{}

\section{Proof of lemmas}\label{app:lemmas}

\setcounter{lemma}{0}
\setcounter{transprops}{0}


\subsection{Proof of Lemma~\ref{comp}}

\begin{lemmaproof}{comp} \label{P-comp}
 If the events of a trace $t$ are events of a program $P$ then so are the events of \\
any completion of $t$.
\[
\all P\dot\all t:Trace\dot\all t^+:ext(t)\dot
 events(t)\subseteq events(P) \,\implies\, events(comp(t^+)) \subseteq events(P)\]
\end{lemmaproof}

\begin{proof}
Let $P$ be a program, $t\mem Trace$ and $t^+\mem ext(t)$.\\
With the definition of $comp$ we have that
$events(comp(t^+)) \subseteq events(t^+)$. \\ 
With the definition of $ext$ it follows that  $\exi tr:\seq Event\dot$ \\
\hspace*{1.9mm} 
              (\refstepcounter{comprops}\label{a}\alph{comprops}) 
                   $ \t1 t\cat tr\mem Trace \land$\vspace*{-0ex} \\
\hspace*{2mm}  
             (\refstepcounter{comprops}\label{b}\alph{comprops})
                $\t1 (\all i < \# tr \dot \exi c:Op\cross Val \dot tr_i=res(c)) \land $ 
                                                                      \vspace*{-0ex}\\
\hspace*{2mm}  
             (\refstepcounter{comprops}\label{c}\alph{comprops})
               $\t1 t^+ = t\cat tr$\vspace*{-0ex}        \\  
From (\ref{c})  we can deduce that 
$events(t^+) = events(t) \uni events(tr)$, \\
and from (\ref{b}) that $\all b:events(tr)\dot
  \exi (op,out):Op\cross Val\dot b=res(op,out)$. \\
With the definition of $Trace$ and (\ref{a}) it follows that
$\exi in:Value\dot inv(op,in)\mem events(t)$.  \\
From this and Axiom~(\ref{events}) we can derive that 
$events(t) \subseteq events(P) \implies events(tr) \subseteq events(P)$, \\
and with $events(t^+) = events(t) \uni events(tr)$ we have that\\
$events(t) \subseteq events(P) \implies events(t^+) \subseteq events(P)$. \\
From this and $events(comp(t^+)) \subseteq events(t^+)$ we can conclude that\\
$events(t) \subseteq events(P) \implies events(comp(t^+)) \subseteq events(P)$.
\end{proof}


\subsection{Proof of Lemma~\ref{Subsetcomp}}

\refstepcounter{theorem}
\refstepcounter{theorem}

Lemma~\ref{Subsetcomp} states that if a trace is allowed by a program then so is any completion that only adds responses for operations whose effects occur in the trace.
The proof of this lemma relies on two sub-results which refer to the construction
of a trace's completion, the addition of responses and the elimination
of pending invocations.

In the following proofs, $S\setminus T$ denotes the set $S$ minus the elements of \vspace{2mm}set $T$.

\noindent{\bf Lemma 3a.}
If a trace $t$ is allowed by $P$ on $M$, then so is any trace formed by adding a response to $t$ when the corresponding effect occurs in $t$.
\[
\all P, M\dot\all res(c):events(P); t:Trace\dot
(\exi i\leq \#t\dot t_i=eff(c)) \land 
\Porder\Subset \order{t} \implies \Porder\Subset \order{t\cat \lseq res(c)\rseq}\]

\begin{proof}
From the definition of $\order{t\cat\lseq res(c)\rseq}$ it follows that
 $\all (a,b):\order{t\cat\lseq res(c)\rseq} \!\!\setminus\!\! \order{t}\dot b=res(c)$.\\
Let $c = (a, out)$ where $inv(a,in)$ and $res(a,out)$ are in $events(P)$, and let $e$ be any event in $events(P)$ such that $(e, res(a,out)) \mem \Porder\!$. For $\Porder\Subset \order{t\cat \lseq res(a,out)\rseq}$ to be true, we need to show that $e$ is in $t$.
\vspace*{1ex}\\
  Case 1:  \parbox[t]{14cm}{$e \neq eff(a,out)$\\ 
   With Axiom~(\ref{enf-res-after}) it follows that 
   $(e, inv(a,in)) \mem\Porder$.\\
   Since $t\cat\lseq res(a,out)\rseq \mem Trace$ we know that 
   $inv(a,in)$ is in $t$ and hence if $\Porder\Subset \order{t}$ (i.e., the second conjunct of the antecedent holds), we can conclude that $e$ is in $t$.}
\vspace*{1ex}\\
  Case 2:  \parbox[t]{14cm}{$e = eff(a,out)$\\
   If $\exi i\leq \#t\dot t_i=eff(c)$ (i.e., the first conjunct of the antecedent holds), then $e$ is in $t$.}
\end{proof}
\vspace*{2ex}

%

Following Lemma~3a., if a trace $t$ is allowed by a program $P$ on memory model $M$, 
the  trace formed by extending $t$ with a sequence of responses whose corresponding effects are in $t$
is also allowed by $P$ on  \vspace{2mm}$M$.

\noindent{\bf Corollary 3.1}
\[\all P, M\dot\all t:Trace\dot \all t\cat tr:ext(t)\dot\\
\t1 (\all i\leq \#tr; c:Op\cross Val\dot tr_i=res(c) \implies \exi j\leq \#t\dot t_j=eff(c)) \land \Porder\Subset \order{t} \implies \Porder\Subset \order{t\cat tr}\]

\noindent{\bf Lemma 3b.}
If, after removing an invocation from a trace, the resulting event
sequence is still a trace, i.e., the invocation was a pending
invocation with neither a response or effect in the trace, then if the original trace is allowed by a program $P$ on
memory model $M$, the resulting trace is allowed by $P$ on $M$.
\[
\all P, M\dot\all inv(c):events(P)\dot\all  t\cat \lseq inv(c)\rseq\cat t':Trace\dot\\
\t1 t\cat t'\mem Trace\land \Porder\Subset \order{t\cat\lseq inv(c)\rseq \cat t'} \implies \Porder\Subset \order{t\cat t'}\]

\begin{proof}
Let $P$ be a program, $M$ be a memory model, $t\cat \lseq inv(c)\rseq\cat t'\mem Trace$, $\Porder\Subset \order{t\cat\lseq inv(c)\rseq \cat t'}$ 
and  $t\cat t'\mem Trace$.\\
We can deduce that $\all a, b : Events \dot (a,b):\order{t\cat\lseq inv(c)\rseq\cat t'}\!\!\setminus\!\!\order{t\cat t'} \implies inv(c) \mem \{a,b\}$. \\
%
Case 1:  \parbox[t]{14cm}{For all $e\mem Event$ where $(e, inv(c)) \mem \Porder$, 
with the definition of $\Subset$,  $(e, inv(c))$ is only enforced by $P_M$ in traces that contain event $inv(c)$.
Since $inv(c) \notin events(t\cat t')$ trace $t \cat t'$ is not affected by the enforced ordering of $(e, inv(c))$.} \\
Case 2: \parbox[t]{14cm}{
For all $e\mem Event$ and $inv(a,in),res(a,out)\mem events(P)$ where $(inv(a,in), e) \mem \Porder$, if $e\neq eff(a,out)$ then with Axiom~(\ref{enf-inv-before}) we also have $(res(a, out), e) \mem \Porder$.\\
Hence if $e \in events(t\cat\lseq inv(a,in)\rseq\cat t')$ then, when $\Porder\Subset \order{t\cat\lseq inv(a,in)\rseq \cat t'}$, we have $e \mem events(t')$ and also that $res(a,out) \mem events(t')$. Hence $t \cat t' \notin Trace$ which contradicts the assumption.

If $e=eff(a,out)$ and $e\in events(t\cat\lseq inv(a,in)\rseq\cat t')$ then, by Axiom~\ref{wellformed}, it is in $t'$. Hence $t \cat t' \notin Trace$ which again contradicts the assumption.\\

On the other hand, if  $e \notin events(t\cat\lseq inv(a,in)\rseq\cat t')$ then $e \notin events(t\cat t')$, 
and the ordering of $(inv(a,in), e)$ is not enforced on trace $t\cat t'$.}\vspace*{2ex}\\
From the above we can conclude that $\Porder \Subset \order{t\cat t'}$.
\end{proof}
\vspace*{2ex}

Following Lemma~3b., if a trace $t$ is allowed by a program $P$ on memory model $M$, the trace formed by removing all pending invocations from $t$ is allowed by $P$ on \vspace{2mm} $M$.

\noindent{\bf Corollary 3.2}
\[\all P, M\dot\all t:Trace\dot \Porder\Subset \order{t} \implies \Porder\Subset \order{comp(t)}\]

\noindent
With  the above two corollaries we can now prove Lemma~\ref{Subsetcomp}.\vspace*{1ex}

\begin{lemmaproof}{Subsetcomp} \label{P-Subsetcomp}
 If a trace $t$ is allowed by a program $P$ on memory model $M$ then so is any completion of $t$.
\[\all P, M\dot\all t:Trace\dot \all t\cat tr:ext(t)\dot\\
~~~~~ (\all i\leq \#tr; c:Op\cross Val\dot tr_i=res(c) \implies \exi j\leq \#t\dot t_j=eff(c)) \land \Porder\Subset \order{t} \implies \Porder\Subset \order{comp(t\cat tr)}\]
\end{lemmaproof}

\begin{proof}
For any $P$, $M$, $t\mem Trace$ and $t\cat tr\mem ext(t)$ where $tr$ only includes responses for operations whose effects occur in $t$, with Corollary~3.2 
we have $\Porder \Subset \order{t\cat tr} \implies \Porder \Subset \order{comp(t\cat tr)}$. 
From this and  Corollary~3.1 
we can deduce that $\Porder \Subset \order{t} \implies \Porder \Subset \order{comp(t\cat tr)}$.
\end{proof}


\subsection{Proof of Lemma~\ref{subset}}

Lemma~\ref{subset} states that the operation order of the completion of a trace is a subset
of the operation order of the original trace. 
The proof of Lemma~\ref{subset} is broken into two steps according to the
two steps of the construction of the completion of a trace.  The first
result states that additional responses that
occur in a completion of a trace do not cause a modification of the
order of operations.  A second result states that the removal of pending
invocation events does not affect the order of the remaining operations. \\

\noindent{\bf Lemma 4a.}
 Adding a response event to a trace $t$ does not affect the operation order $\supclass_{t}$.
 \[
\all c:Op\cross Val \dot\all t\cat \lseq res(c) \rseq :Trace\dot 
     \,\,\supclass_{t}\ =\ \supclass_{t\cat\lseq res(c)\rseq}\]

 \begin{proof}
Since $\supclass_t$ only includes pairs $(res(c),inv(d))$, i.e., where $res(c)$ occurs in trace $t$ before $inv(d)$, adding a response at the end of $t$ does not change $\supclass_t$. 
\end{proof}\\

As a consequence of Lemma 4a  we can conclude that 
extending a trace $t$ with a sequence of response events 
does not affect \vspace*{2mm}$\supclass_{t}$.

\noindent{\bf Corollary 4.1}
\[\all t:Trace\dot \all t^+:ext(t)\dot \ \supclass_{t}\ =\ \supclass_{t^+}\]

\noindent{\bf Lemma 4b.}
If, after removing an invocation from a trace, the resulting event sequence is still a trace, 
i.e., the invocation was a pending invocation, then its $\supclass$ order is a subset of that of the original trace.
 \[
\all t\cat \lseq inv(c)\rseq \cat t':Trace\dot t\cat t'\mem Trace \implies \ 
      \supclass_{t\cat t'}\ \subseteq \ \supclass_{t\cat\lseq inv(c)\rseq \cat t'}\]

 \begin{proof}
For any $t \cat\lseq inv(c)\rseq \cat t' \mem Trace$, assume $t\cat t'\mem Trace$.\\
From the definition of $Trace$ we have
  $inv(c) \nmem events(t) \land inv(c) \nmem events(t')$. \\
Then from the definition of $\supclass_{t \cat t'}$ it follows that 
$\supclass_{t \cat t'}= \supclass_{t \cat \lseq inv(c)\rseq \cat t'}\!\setminus\, \{(a,b):<_{t \cat \lseq inv(c)\rseq \cat t'}| b=inv(c)\}$.\\
Hence, we can deduce that
  $\ \supclass_{t\cat t'}\ \subseteq \ \supclass_{t\cat\lseq inv(c)\rseq \cat t'}$. 
\end{proof}\\

Following Lemma~4b., removing all pending invocations from a trace results in a trace 
whose $\supclass$ order is a subset of that of the original \vspace*{2mm}trace.

\noindent{\bf Corollary 4.2}
\[\all t:Trace\dot\ \supclass_{comp(t)}\ \subseteq\ \supclass_{t}\]

The two corollaries provide us now with a succinct proof of Lemma~\ref{subset}.\\

\begin{lemmaproof}{subset}  
\label{P-subset}
The $\supclass$ order of a completion of a trace $t$ is a subset of that of $t$.
\[\all t:Trace\dot \all t^+:ext(t)\dot\ \supclass_{comp(t^+)}\ \subseteq\ \supclass_{t}\]
\end{lemmaproof}
\vspace*{-1.8ex}

\begin{proof}
Let $t\mem Trace$ and $t^+\mem ext(t)$. With Corollary 4.2 
it follows that $\ \supclass_{comp(t^+)}\ \subseteq \ \supclass_{t^+}$ 
and together with Corollary 4.1  
we can conclude that
$\ \supclass_{comp(t^+)}\ \subseteq \ \supclass_{t}$.
\end{proof}


\section{Transformation function proofs}\label{app:sound}

\setcounter{transprops}{0}
According to case (T4) in Definition~\ref{transdef} the order of program events in $t'$ is the same as in 
$comp(t^+)$.
\[ \forall a,b : \PEv \dot 
       (a,b)\in \order{comp(t^+)} \implies (a,b) \in \order{t'} & (\refstepcounter{transprops}\roman{transprops}\label{trans-ps})
\]

Operation events may be reordered in $t'$ compared to $comp(t^+)$
unless their order in $h'$ is the same as that in $comp(t^+)$:
\begin{itemize}
\item[-] according to case (T1) an invocation will be placed at the same
  place in $t'$ if $comp(t^+)$ and $h'$ coincide on that event.
  \item[-] according to case (T2), invocations in $comp(t^+)$ that do not match the
  head of $h'$ will be placed into the remainder sequence $r$, whose content will be added later to $t'$.
  \item[-] according to step (T5), an invocation which has been placed
  into the remainder sequence $r$ can be placed in $t'$ when it matches the
  event at the head of $h'$. Its occurrence in $t'$ will be later than in $comp(t^+)$.
  \item[-] according to case (T3) the occurrence of responses and effects in $comp(t^+)$ will be discarded and instead these events will be placed directly after their invocation event ((T1) and (T5)) to mimic atomicity of the abstract operation.
  \item[-] note that by (T4) program events will only be placed in $t'$  if 
    invocations at the head of $h'$ and $r$ do not match,
    i.e., placing of object events has precedence over placing of program events. 
\end{itemize}
It follows that, with respect to program events, invocations will either
stay where they are or will be moved to a later place in $t'$. It is never the
case that an invocation will be placed at an earlier place in $t'$
(invocations only ever move to the right) w.r.t. program events.
\[ \forall x : \PEv, c:Op\cross Val \dot 
             (x, inv(c)) \in \order{comp(t^+)} \implies (x, inv(c)) \in \order{t'}
           & (\refstepcounter{transprops}\roman{transprops}\label{trans-inv})
\]
With respect to program events, responses and operation effects will
always be placed earlier in $t'$ but never later than their placement in
$comp(t^+)$. Responses and operation effects only ever move to the left
w.r.t. program events. If the invocation stays where it is in the order,
then its response and effect will move to an earlier point (due to
(T1) and (T5)). If the invocation event is placed at a later point it will never
be placed after a program event that followed its response event (due to the
fact that the placement of object events has precedence over the placement of
program events). 
\[ \forall x : \PEv, c:Op\cross Val \dot 
             (res(c), x) \in \order{comp(t^+)} \implies (res(c), x) \in \order{t'}
           & (\refstepcounter{transprops}\label{trans-res}\roman{transprops})
\]
Due to (T1) and (T5) operation effects will be placed at the earliest point in $t'$ (i.e., directly 
before the response event). Hence, with (\ref{trans-res}) it follows that 
\[ \forall x : \PEv, c:Op\cross Val \dot 
             (\obs(c), x) \in \order{comp(t^+)} \implies (\obs(c), x) \in \order{t'}
           & (\refstepcounter{transprops}\label{trans-obs}\roman{transprops})
\]


The above properties of $trans$, (\ref{trans-ps})--(\ref{trans-obs}), and Axioms
(\ref{enf-inv-before})--(\ref{enf-order-obs}) of $\!\!\Porder\!\!$  (on page \pageref{enf-inv-before}) 
enable us to prove the four cases which show that $P_M$ allows the constructed
trace $t'$, i.e., $\Porder \Subset \order{t'}$, from which we can conclude that $t' \in \semM{P}$.

\begin{itemize}
\item[a.)]~ $\all a,b:\PEv\dot (a, b)\in\Porder \land b\in events(t')\implies 
                                    (a,b)\in\order{t'}$\\
\begin{proof}
    Since $\Porder \Subset\ <_{comp(t^+)}$ (\ref{sp_3}) \\
   $\all a, b:\PEv\dot (a, b)\in\Porder \land b\in events(comp(t^+))\implies (a, b)\in\order{comp(t^+)}$\\
   with (\ref{trans-ps}) we have\\
 $\all a, b:\PEv\dot (a, b)\in\Porder \land b\in events(comp(t^+))\implies (a, b)\in\order{t'}$\\
and (\ref{sp_6a})\\
   $\all a, b:\PEv\dot (a, b)\in\Porder \land b\in events(t')\implies (a, b)\in\order{t'}$
\end{proof}\vspace*{1ex}

   Hence the order of program events that is enforced by $P_M$ is maintained by $t'$.\\
\item[b.)] ~$\all p :\PEv; e:\ObjEv\dot
                        (p,e)\in\Porder \land e\in events(t')\implies (p,e)\in\order{t'} $\\
\begin{proof}
$Case ~b1: e$ is an invocation event\\
    Since $\Porder \Subset\ <_{comp(t^+)}$ (\ref{sp_3}) \\
    $ \all p :\PEv; c:(Op\cross Val) \dot\\
     \hspace*{1cm}
             (p,inv(c))\in\Porder \land inv(c)\in events(comp(t^+))\implies (p,inv(c))\in\order{comp(t^+)}$\\
    with (\ref{trans-inv}) we have\\
$ \all p :\PEv; c:(Op\cross Val) \dot
               (p,inv(c))\in\Porder \land inv(c)\in events(comp(t^+))\implies (p,inv(c))\in\order{t'}$\\
and (\ref{sp_5})\\
     $ \all p :\PEv; c:(Op\cross Val) \dot
               (p,inv(c))\in\Porder \land inv(c)\in events(t')\implies (p,inv(c))\in\order{t'}\\
    \hspace*{1cm}$

  $Case ~b2: e$ is a response event\\
     Let $inv(a,in)$ and $res(a,out)$ be in $events(P)$, and assume that $res(a,out)\mem events(t')$.   \\
     Since $t'\mem Trace$  (\ref{sp_8}), $inv(a,in)\mem events(t')$ (Axiom \ref{wellformed}), and from $(Case~b1)$  \\
     $(p,inv(a,in))\in\Porder \implies (p,inv(a,in))\in\order{t'}$\\
     and since $inv(a,in)$ must occur in $t'$ before $res(a,out)$ (Axiom \ref{wellformed})\\
     $(p,inv(a,in))\in\Porder \implies (p,res(a,out))\in\order{t'}$.\\
     Hence, with Axiom \ref{enf-res-after} we have\\
     $(p,res(a,out))\in\Porder \implies (p,res(a,out))\in\order{t'}$\\
     and therefore\\
      $ \all p :\PEv; c:(Op\cross Val) \dot
               (p,res(c))\in\Porder \land res(c)\in events(t')\implies (p,res(c))\in\order{t'}\\
   $
     
$Case ~b3: e$ is an effect event\\
   Following the same line of reasoning as in $Case~b2$ (with Axiom~\ref{enf-obs-after-ps} in place of Axiom \ref{enf-res-after}) we have\\
    $\all p :\PEv; c: (Op\cross Val)\} \dot
           (p,\obs(c))\in\Porder \land \obs(c)\in events(t')\implies (p,\obs(c))\in\order{t'}\\
      $

Hence, with $(Case~b1)$, $(Case~b2)$ and $(Case~b3)$ it follows that
the order of a program event followed by an object event that is enforced by $P_M$ is maintained in $t'$,
i.e., \\
$\all p :\PEv; e:\ObjEv\dot
                 (p,e)\in\Porder \land e\in events(t')\implies (p,e)\in\order{t'} $
\end{proof}
\item[c.)]~~ $\all p :\PEv; e:\ObjEv\dot
                 (e, p)\in\Porder \land p\in events(t')\implies (e, p)\in\order{t'}$\\

\begin{proof}
$Case ~c1: e$ is an invocation event\\
   Since $\Porder \Subset\ <_{comp(t^+)}$ (\ref{sp_3}) \\
   $ \all p :\PEv; res(a,out):events(P) \dot\\
     \hspace*{1cm} (res(a,out), p)\in\Porder \land p\in events(comp(t^+))\implies (res(a,out), p)\in\order{comp(t^+)}$\\
   and with  Axiom (\ref{enf-inv-before}) \\
   $ \all p :\PEv; inv(a,in),res(a,out):events(P) \dot\\
     \hspace*{1cm} (inv(a,in), p)\in\Porder \land p\in events(comp(t^+))\implies (res(a,out), p)\in\order{comp(t^+)}$\\
     and with (\ref{sp_6a})\\
      $ \all p :\PEv; inv(a,in),res(a,out):events(P) \dot\\
     \hspace*{1cm} (inv(a,in), p)\in\Porder \land p\in events(t')\implies (res(a,out), p)\in\order{comp(t^+)}$.\\
     With (\ref{trans-res}) we have\\
     $ \all p :\PEv; inv(a,in),res(a,out):events(P) \dot\\
 \hspace*{1cm}(inv(a,in), p)\in\Porder \land p\in events(t')\implies (res(a,out), p)\in\order{t'}$\\
     and with Axiom (\ref{wellformed})\\
   $ \all p :\PEv; inv(a,in):events(P) \dot\\
\hspace*{1cm} (inv(a,in), p)\in\Porder \land p\in events(t')\implies (inv(a,in), p)\in\order{t'}\\
     $ 
     
    $Case ~c2: e$ is a response event\\
    Since $\Porder \Subset\ <_{comp(t^+)}$ (\ref{sp_3}) \\
    $\all p :\PEv; c:Op\cross Val \dot      (res(c), p)\in\Porder \land p\in events(comp(t^+))\implies (res(c), p)\in\order{comp(t^+)}$\\
    with (\ref{trans-res}) we have\\
    $\all p :\PEv; c:Op\cross Val\dot
      (res(c), p)\in\Porder \land p\in events(comp(t^+))\implies   (res(c), p)\in\order{t'}$  \\
    and with (\ref{sp_6a}) \\
    $ \all p :\PEv; c:Op\cross Val\dot
       (res(c), p)\in\Porder \land p \in events(t')\implies (res(c), p)\in\order{t'}\\
      $
      
$Case ~c3: e$ is an effect event\\
Following the same line of reasoning as in $Case~c2$ (with (\ref{trans-obs}) in place of (\ref{trans-res})) we have\\
    $ \all p :\PEv; c:Op\cross Val\dot
       (\obs(c), p)\in\Porder \land p \in events(t')\implies (\obs(c), p)\in\order{t'}\\
      $
      
      Hence, with $(Case~c1)$, $(Case~c2)$ and $(Case~c3)$ it follows that
the order of an object event followed by a program event that is enforced by $P_M$ is maintained in $t'$,
i.e.,\\
$\all p :\PEv; e:\ObjEv\dot
                  \hspace*{0.2cm} (e,p)\in\Porder \land p\in events(t')\implies (e,p)\in\order{t'}$
\end{proof}\vspace*{1ex}
\\

\item[d.)]~~ $\all a,b:\ObjEv\dot (a, b)\in\Porder \land b\in events(t')\implies (a,b)\in\order{t'}$\\
\begin{proof}
$Case ~d1: a$ and $b$ are events of the same operation\\
If the object events refer to the same operation, the reasoning is simply based on the fact that (T5) ensures that the events occur in $t'$ in the order (i) invocation before effect, and (ii) effect before response, and no program/memory model combination can enforce an order contrary to this.  (i) is common to all programs on all memory models, and (ii) can occur on any program on any memory model (by adding a fence to the operation if the memory model is not SC; under SC the effect always occurs immediately before the response).\\  

$Case ~d2: a$ and $b$ are events of different operations\\
If the object events refer to two different operations, we first prove that if an operation must occur before another in $P$ on $M$ then this is also the case in $t'$.

  Since $\Porder \Subset\ <_{comp(t^+)}$ (\ref{sp_3}) \\
$ \all c, d : (Op \cross Val) \dot
    (res(c), inv(d))\in\Porder \land inv(d) \in events(comp(t^+))\implies (res(c), inv(d))\in\order{comp(t^+)}$\\
  with $\resIR{comp(t^+)} = comp(h^+)$ (\ref{sp_2a}) and the definition of $ \supclass_{t}$ 
  (on page \pageref{supclass_def}) we have \\
$ \all c, d : (Op \cross Val)\dot  (res(c), inv(d))\in\Porder \land inv(d) \in events(comp(t^+))\implies
  (res(c), inv(d)) \in \supclass_{comp(h^+)}$\\
  with $\supclass_{comp(h^+)}\ \subseteq\ \supclass_{h'}$ (\ref{sp_1}) and $events(\resIR{t'})=events(\resIR{comp(t^+)})$ (\ref{sp_5})\\
$ \all c, d : (Op \cross Val) \dot (res(c), inv(d))\in\Porder \land inv(d) \in events(t')\implies
       (res(c), inv(d)) \in \supclass_{h'}$\\
  with $\resIR{t'} = h'$ (\ref{sp_5}) and the definition of $ \supclass_{t}$\\
$ \all c, d : (Op \cross Val) \dot (res(c), inv(d))\in\Porder \land inv(d) \in events(t')\implies
        (res(c), inv(d)) \in \order{t'} ~\hfill(*)$\\ 
 
Using this result, we then show that if the events of the operations must occur in a specific order in $P$ on $M$ then this is also the case in $t'$.

Let $a\mem \{inv(c,in_c), res(c,out_c), \obs(c,out_c)\}$ and $b\mem \{inv(d,in_d),res(d,out_d),\obs(d,out_d)\}$ where $c\neq d$, and assume $(a,b) \mem \Porder$. Then from Lemma~\ref{porderLemma} we have $(res(c,out_c), inv(d,in_d)) \mem \Porder$.\\
Hence, from $(*)$ we have\\
$(a, b)\in\Porder \land inv(d,in_d) \in events(t')\implies (res(c,out_c), inv(d,in_d)) \in \order{t'}$\\ 
Since (T1) and (T5) ensure that if $inv(d)$ occurs in $t'$ so do $eff(d,out_d)$ and $res(d,out_d)$, we have\\
$(a, b)\in\Porder \land b \in events(t')\implies (res(c,out_c), inv(d,in_d)) \in \order{t'}$\\ 
and since (T1) and (T5) also ensure that the invocation, effect and response of an operation occur together\\
$(a, b)\in\Porder \land b \in events(t')\implies (a, b) \in \order{t'}$ \\ 

 
From $(Case~d1)$ and $(Case~d2)$ it follows that\\
$\all a,b:\ObjEv\dot (a, b)\in\Porder \land b\in events(t')\implies (a,b)\in\order{t'}$
\end{proof}

\end{itemize}



\section{Completion of matching traces}\label{app:complete}

\begin{lemmaproof}{matchingTraces}
\label{P_cp_2}
 For an object implementation $C$ which is an object refinement of a specification $A$, 
and any client program $P$ that records every invocation and response of an operation without delay, we have
\[\all t : \semM{{P}[C]} \dot (\exi t':\semM{{P}[A]}\dot \obsT{t}=\obsT{t'})\implies \\
    \t2 \exi t'':\semM{{P}[A]} \dot \obsT{t}=\obsT{t''}   \land \exi t^+:ext(t) \dot \resIR{comp(t^+)}\sim \resIR{t''}    \]
\end{lemmaproof}

\begin{proof} For any client program ${P}$ that records every invocation and response of an operation without delay it holds that \emph{matching} traces of $\semM{{P}[A]}$ and $\semM{{P}[C]}$, i.e., those traces for which the sequence of observable steps is the same, will differ by at most one invocation/response pair per thread (since after such a pair there would be a further observable event before the next invocation). Hence, we consider the following 6 cases:

\begin{enumerate}
\item $t$ and $t'$ have exactly the same sequences of invocations and responses per thread. In this case, the lemma is satisfied by choosing $t''=t'$ and $t^+=t$.
\item $t'$ has extra invocation/response pairs (note that abstract traces cannot be extended by just an invocation (see Axiom~(\ref{seqSpec})). In this case, we choose $t''$ by removing the invocation/response pairs. This is always possible since the $\semM{P[A]}$ is prefix-closed (and there is at most one invocation/response pair on each thread). Then with 
$t^+=t$, the lemma is satisfied. 
\item $t$ has extra invocation/response pairs. Since all responses are recorded by $P$, we know that there is an extension of $t$ in $\semM{P[C]}$ with observable program steps recording each of the extra pairs. From the antecedent, there must be a matching trace in $\sem{P[A]}$ and hence we know that there exists a $t''$ which extends $t'$ with the invocation/response pairs of $t$. Given this $t''$, the lemma is satisfied with $t^+=t$.
\item $t$ and $t'$ both have invocation/response pairs, but they are different. We use the prefix of $t'$ which does not have the invocation/response pairs (as in case~2), and extend it with the invocation/response pairs of $t$ (as in case~3) to get $t''$. Then the lemma is satisfied with $t^+=t$.
\item $t$ has one or more extra invocations. In this case, for each invocation with an effect we add a response in $t^+$ and we extend $t'$ with the matching invocation/response pairs (as in case~4) to get $t''$. The remaining invocations will be removed by the function $comp$.
\item $t$ has one or more extra invocations, and $t'$ has extra invocation/response pairs. In this case, we use the prefix of $t'$ which does not have the extra invocation/response pairs (as in case~2), and extend $t$ and this prefix of $t'$ as in case~5.
\end{enumerate}
~\end{proof}

\end{document}